\documentclass[11pt]{article} 

\usepackage[utf8]{inputenc} 
\usepackage[toc, page]{appendix}
\usepackage{amsfonts}
\usepackage{bm}
\usepackage{amsmath}
\usepackage{bbm}
\numberwithin{equation}{section}
\numberwithin{figure}{section}
\numberwithin{table}{section}

\usepackage[ruled,vlined, linesnumbered]{algorithm2e}

\usepackage[total={6.5in, 8in}]{geometry} 
\usepackage{graphicx} 
\usepackage{float}
\usepackage{caption}
\usepackage{subcaption}
\usepackage{rotating}

\usepackage{booktabs} 
\usepackage{array} 
\usepackage{paralist} 
\usepackage{verbatim} 
\usepackage{hyperref}
\hypersetup{pdftex,colorlinks=true,allcolors=blue}
\usepackage{hypcap}
\usepackage{amsthm}

\usepackage{fancyhdr} 
\usepackage{sectsty}
\allsectionsfont{\sffamily\mdseries\upshape} 

\usepackage[nottoc,notlof,notlot]{tocbibind} 
\usepackage[titles,subfigure]{tocloft} 

\usepackage[font=small,labelfont=bf,tableposition=top]{caption}

\usepackage{authblk}
\usepackage[export]{adjustbox}
\usepackage{mathtools}

\DeclarePairedDelimiter\floor{\lfloor}{\rfloor}

\theoremstyle{definition}
\newtheorem{definition}{Definition}[section]
\newtheorem{remark}[definition]{Remark}

\newcommand{\PreserveBackslash}[1]{\let\temp=\\#1\let\\=\temp}
\newcolumntype{C}[1]{>{\PreserveBackslash\centering}p{#1}}
\newcolumntype{R}[1]{>{\PreserveBackslash\raggedleft}p{#1}}
\newcolumntype{L}[1]{>{\PreserveBackslash\raggedright}p{#1}}

\DeclareCaptionLabelFormat{andtable}{#1~#2  \&  \tablename~\thetable}

\providecommand{\keywords}[1]
{
  \small	 
  \textbf{\textit{Keywords---}} #1
}

\title{Deep calibration of the quadratic rough Heston model \thanks{
  This work benefits from the financial support of the Chaires Machine Learning \& Systematic Methods, 
  Analytics and Models for Regulation, and the ERC Grant 679836 Staqamof. The authors would like to thank Paul Gassiat, 
  Jim Gatheral, Julien Guyon, Paul Jusselin, Marouane Anane and Alexandre Davroux for very useful comments.
}}

\author{
  Mathieu Rosenbaum $^1$
  \qquad 
  Jianfei Zhang $^{1, 2}$
}%

\date{%
    $^1$ \footnotesize Ecole Polytechnique, CMAP, 91128 Palaiseau Cedex, France \\%
    $^2$ \footnotesize Exoduspoint Capital Management, 32 Boulevard Haussmann, 75009 Paris, France \\[2ex]%
\today}
\begin{document}
\maketitle

\begin{abstract}
  The quadratic rough Heston model provides a natural way to encode Zumbach effect in the rough volatility paradigm.
  We apply multi-factor approximation and use deep learning methods to build an efficient
  calibration procedure for this model.
  We show that the model is able to reproduce very well both SPX and VIX implied volatilities.
  We typically obtain VIX option prices within the bid-ask spread and an excellent fit of the SPX at-the-money skew.
  Moreover, we also explain how to use the trained neural networks for hedging with instantaneous computation of
  hedging quantities. 
\end{abstract}
\keywords{Quadratic rough Heston, multi-factor approximation, SPX smile, VIX smile, deep learning, joint calibration, hedging}

\section{Introduction}
The rough volatility paradigm introduced in \cite{gatheral2018volatility} is now widely accepted, both by practitioners
and academics. On the macroscopic side, rough volatility models can fit with remarkable accuracy the shape of implied
volatility smiles and at-the-money skew curves. They also reproduce amazingly well stylized facts of realized
volatilities, see for example \cite{bayer2016pricing, bennedsen2016decoupling, el2019roughening,gatheral2018volatility, livieri2018rough}.
On the microstructural side, it is shown in \cite{dandapani2021quadratic, el2018microstructural,jaisson2016rough} that the rough Heston
model introduced and developed in \cite{el2018perfect, el2019characteristic} naturally emerges from agents behaviors
at the microstructural scale. 
\vskip 0.2in
\noindent Nevertheless, one stylized fact of financial time series that is not reflected in the rough Heston model is the
feedback effect of past price trends on future volatility, which is discussed by Zumbach in \cite{zumbach2010volatility}.
Super-Heston rough volatility models introduced in \cite{dandapani2021quadratic} fill this gap by
considering quadratic Hawkes processes from microstructural level, and showing that the Zumbach effect remains explicit
in the limiting models. As a particular example of super-Heston rough volatility models, the authors in \cite{gatheral2020quadratic}
propose the quadratic rough Heston model, and show its promising ability to calibrate jointly SPX smiles and VIX smiles,
where other continuous-time models have been struggling for a long time \cite{guyon2020joint}.

\vskip 0.2in
\noindent
The VIX index is in fact by definition a derivative of the SPX index $S$, which can be represented as 
\begin{equation}
  \text{VIX}_t = \sqrt{-2\mathbb{E}[\log(S_{t+\Delta}/S_t)|\mathcal{F}_t]} \times 100 \, ,
  \label{eq:vix}
\end{equation}
where $\Delta=30$ days and $\mathbb{E}$ is the risk-neutral expectation. Consequently, VIX options are also derivatives of SPX.
Finding a model which jointly calibrates the prices of SPX and VIX options is known to be very challenging, especially
for short maturities. As indicated in \cite{guyon2020joint}, ``the very negative skew of short-term SPX options, 
which in continuous models implies a very large volatility of volatility, seems inconsistent with the comparatively
low levels of VIX implied volatilities". Through numerical examples, the authors in \cite{gatheral2020quadratic}
show the relevance of the quadratic rough Heston model in terms of pricing simultaneously SPX and VIX options.
In this paper, in the spirit of \cite{abi2019multifactor}, we propose a multi-factor approximated version of this model
and an associated efficient calibration procedure. 

\vskip 0.2in
\noindent 
Under the rough Heston model, the characteristic function of the log-price
has semi-closed form formula, and thus fast numerical pricing methods can be designed, see \cite{el2019characteristic}. 
However, pricing in the quadratic
rough Heston model is more intricate. The multi-factor approximation method for the rough kernel function developed 
in \cite{abi2019multifactor} makes rough volatility models Markovian in high dimension. Thus Monte-Carlo simulations
become more feasible in practice. Still, in our case, model calibration remains a difficult task. 
Inspired by recent works about applications of deep learning in quantitative finance, see for example
\cite{bayer2019deep, hernandez2016model, horvath2021deep}, we use deep neural networks to speed up model calibration.
The effectiveness of the calibrated model for fitting jointly SPX and VIX smiles is illustrated through numerical
experiments. Interestingly, under our model, the trained networks also allow us to hedge options with instantaneous computation
of hedging quantities.

\vskip 0.2in
\noindent The paper is organized as follows. In Section \ref{section:model}, we give the definition of our model and
introduce the approximation method. In Section \ref{section:dl}, we develop the model calibration with deep neural
networks. Validity of the methods is tested both on simulated data and market data. Finally in Section \ref{section:hedging}, 
we show how to perform hedging in the model with neural networks through some toy examples.

\section{The quadratic rough Heston model and its multi-factor approximation}
\label{section:model}
The quadratic rough Heston model, proposed in \cite{gatheral2020quadratic}, for the price of an asset $S$(here the SPX)
and its spot variance $V$ under risk-neutral measure is
\begin{equation}
    dS_t = S_t\sqrt{V_t}dW_t,  \qquad V_t = a(Z_t - b)^2 + c\, ,
    \label{eq:qudratic_v}
    \footnote{To ensure the martingale property of $S_t$, we can in fact use $V_t=a\phi(Z_t-b)+c$, where $\phi$ is defined as
      $$
        \phi(x) = \begin{cases}
            x^2 & \quad \text{if } x<x^{*} \\ 
            (x^*)^2 &  \quad \text{otherwise}
        \end{cases}
      $$
      with $x^*$ sufficiently large. In this paper, for ease of notation we keep writing the square function.}
\end{equation}
where $W$ is a Brownian motion, $a, b, c$ are all positive constants and $Z_t$ is defined as
\begin{equation}
    Z_t = \int_0^t \lambda\frac{(t-s)^{\alpha - 1}}{\Gamma(\alpha)}\big(\theta_0(s) - Z_s\big)ds + 
    \int_0^t\eta\frac{(t-s)^{\alpha-1}}{\Gamma(\alpha)}\sqrt{V_s}dW_s  \, ,
  \label{eq:Z_t}
\end{equation}
for $t\in[0, T]$. Here $T$ is a positive time horizon, $\alpha\in(1/2, 1)$, $\lambda>0$, $\eta>0$ and $\theta_0(\cdot)$ is
a deterministic function. $Z_t$ is driven by the returns through $\sqrt{V_t}dW_t = dS_t/S_t$.
Then the square in $V_t$ can be understood as a natural way to encode the so called \textit{strong Zumbach effect}, which means that
the conditional law of future volatility depends not only on path volatility trajectory but also on past returns. Note that
in this case we have a pure-feedback model as $S_t$ and $V_t$ are driven by the same Brownian motion, see \cite{dandapani2021quadratic} for
more details on the derivation of this type of models.
We will see in Section \ref{section:hedging} that this setting
enables us to hedge perfectly European options with SPX only. We recall the parameter interpretation given in \cite{gatheral2020quadratic}:
\begin{itemize}
  \item $a$ stands for the strength of the feedback effect on volatility. 
  \item $b$ encodes the asymmetry of the feedback effect. It reflects the empirical fact that negative price returns
        can lead to volatility spikes, while it is less pronounced for positive returns. 
  \item $c$ is the base level of variance, independent from past prices information.
\end{itemize}
\vskip 0.2in
\noindent
It is shown in \cite{el2019characteristic} that under the rough Heston model the volatility trajectories have almost surely 
H$\ddot{\text{o}}$lder regularity $\alpha - 1/2 -\varepsilon$, for any $\varepsilon > 0$. This actually recalls the observation
in \cite{gatheral2018volatility} that the dynamic of log-volatility is similar to that of a fractional Brownian motion with Hurst
parameter of order 0.1. Similarly the fractional kernel
$K(t) = \frac{t^{\alpha-1}}{\Gamma(\alpha)}$ in (\ref{eq:Z_t}) enables us to generate rough
volatility dynamics, which is highly desirable as explained in the introduction. However, it makes the quadratic rough
Heston model non-Markovian and non-semimartingale, and thus difficult to simulate efficiently. In this paper, we apply
the multi-factor approximation proposed in \cite{abi2019multifactor} to do so. The key idea is to write the fractional kernel
$K(t)$ as the Laplace transform of a positive measure $\mu$
$$
  K(t) = \int_0^\infty e^{-\gamma t}\mu(d\gamma), \quad \mu(d\gamma) = \frac{\gamma^{-\alpha}}{\Gamma(\alpha)\Gamma(1-\alpha)}d\gamma \, .
$$
Then we approximate $\mu$ by a finite sum of Dirac measures $\mu^n = \sum_{i=1}^nc_i^n\delta_{\gamma_i^n}$ with
positive weights $(c_i^n)_{i=1,\cdots,n}$ and discount coefficients $(\gamma_i^n)_{i=1,\cdots,n}$, with $n \geq 1$.
This gives us the approximated kernel function
$$
  K^n(t) = \sum_{i=1}^nc_i^ne^{-\gamma^n_i t}, \quad n\geq 1 \, .
$$
A well-chosen parametrization of the $2n$ parameters $(c^n_i, \gamma^n_i)_{i=1,\cdots,n}$ in terms of $\alpha$
can make $K^n(t)$ converge to $K(t)$ in the $L^2$ sense as $n$ goes to infinity, and the multi-factor approximation
models behave closely to their counterparts in the rough volatility paradigm, see \cite{abi2019lifting, abi2019multifactor}
for more details. We recall in Appendix \ref{ap:kernel_aprox} the parametrization method proposed in \cite{abi2019lifting}. 
Then given the time horizon $T$ and $n$, $(c^n_i)_{i=1,\cdots,n}$ and $(\gamma^n_i)_{i=1,\cdots,n}$
are just deterministic functions of $\alpha$, and therefore not free parameters to calibrate.      
We can give the following multi-factor approximation of the quadratic rough Heston model:
\begin{align}
  dS^n_t &= S^n_t\sqrt{V^n_t}dW_t, \quad V^n_t = a(Z^n_t - b)^2 + c \, \label{eq:V_n} \, ,\\
  Z^n_t &= \sum_{i=1}^nc^n_iZ^{n, i}_t\, \label{eq:Z_n} \, ,\\
  dZ^{n,i}_t &= (-\gamma^n_iZ^{n,i}_t - \lambda Z^n_t)dt + \eta\sqrt{V^n_t}dW_t, \qquad Z^{n,i}_0 = z^i_0 \, ,
\label{eq:Z_factor}
\end{align}
with $(z^i_0)_{i=1,\cdots,n}$ some constants. Contrary to the case of the rough Heston model, $\theta(t)$ cannot be easily
written as a functional of the forward variance curve in the quadratic rough Heston model. Then instead of making the 
factors $(Z^{n,i})_{i=1,\cdots,n}$ starting from 0 and taking $Z_t^n = \lambda\sum_{i=1}^nc_i^n\int_0^te^{-\gamma_i^n(t-s)}\theta(s)ds + \sum_{i=1}^nc^n_iZ_t^{n,i}$,
as the authors do in \cite{abi2019lifting, abi2019multifactor}, here we discard $\theta(\cdot)$ in Equation (\ref{eq:Z_n})
and consider the starting values of factors $(z^i_0)_{i=1,\cdots,n}$ as free parameters to calibrate from market data.
This setting allows $Z_0$ and $V_0$ to adapt to market conditions and also encodes various possibilities for
the ``term-structure" of $Z^{n}_t$. To see this, given a solution for 
(\ref{eq:V_n})-(\ref{eq:Z_factor}), (\ref{eq:Z_factor}) can be rewritten as
\begin{equation}
   Z^{n,i}_t = z_0^ie^{-\gamma^n_i t} + \int_0^te^{-\gamma^n_i(t-s)}\big(-\lambda Z^n_sds + \eta\sqrt{V^n_s}dW_s\big) \, .
    \label{eq:Z_factor_int}
  \end{equation}
Then from (\ref{eq:Z_n}) we can get
\begin{equation}
  Z^n_t = g^n(t) + \int_0^tK^n(t-s)\big(b(Z^n_s)ds + \sigma(Z^n_s)dW_s\big)\, , 
  \label{eq:Z_multi}
\end{equation}
with $g^n(t) = \sum_{i=1}^nz^i_0c^n_ie^{-\gamma_i^nt}$, $b(Z^n_t) = -\lambda Z^n_t$ and $\sigma(Z^n_t) = \eta\sqrt{a(Z^n_t - b)^2+c}$. 
By taking expectation on both sides of (\ref{eq:Z_multi}), we get 
$$
  \mathbb{E}[Z_t^n] + \lambda\sum_{i=1}^nc_i^n\int_0^te^{-\gamma_i(t-s)}\mathbb{E}[Z_s^n]ds = \sum_{i=1}^nz_0^ic_i^ne^{-\gamma^n_it} \, .
$$
Thus we can see that the $(z^i_0)_{i=1,\cdots, n}$ allows us to encode initial ``term-structure" of $Z_t^n$. 
Therefore, it can be understood as an analogy of $\theta(t)$ for the variance process in the rough Heston model.
Besides, we will see in Section \ref{section:hedging} that this setting allows us to hedge options perfectly with only SPX.

\vskip 0.2in
\noindent
By virtue of Proposition B.3 in \cite{abi2019multifactor}, for given $n$, Equations (\ref{eq:Z_multi}) and equivalently (\ref{eq:Z_factor_int}) admit
a unique strong solution, since $g^n(t)$ is H$\ddot{\text{o}}$lder continuous, $b(\cdot)$ and $\sigma(\cdot)$ have linear growth, and $K^n$ is 
continuously differentiable admitting a resolvent of the first kind. We stress again the fact that Model (\ref{eq:V_n}-\ref{eq:Z_factor}) 
does not bring new parameters to calibrate compared to the quadratic rough Heston model defined in (\ref{eq:qudratic_v}-\ref{eq:Z_t}), 
with the idea of the correspondance between $\theta(t)$ and $(z^i_0)_{i=1,\cdots,n}$. The fractional kernel in the rough volatility paradigm 
helps us to build the factors $(Z^{n,i})_{i=1,\cdots,n}$. Factors with large discount coefficient $\gamma^n_i$ can mimic roughness and 
account for short timescales, while ones with small $\gamma^n_i$ capture information from longer timescales. The quantity $Z^n$ aggregates 
these factors and therefore encodes the multi-timescales nature of volatility processes, which is discussed for example in 
\cite{fouque2011multiscale, gatheral2018volatility}.

\begin{remark}
  With multi-factor approximation, Model (\ref{eq:V_n}-\ref{eq:Z_factor}) is Markovian with a state vector of dimension $n+1$ given by $\mathbf{X}^n_t :=(S^n_t, Z^{n,1}_t, \cdots, Z^{n,n}_t)$. 
  Hence the price of SPX options at time $t$ is fully determined by $\mathbf{X}^n_t$.
  \label{re:markov_model}
\end{remark}

\noindent 
As discussed in Appendix \ref{ap:kernel_aprox}, we choose $n=10$ in our numerical experiments. To simplify notations, we discard the label
$n$ in the following, and let $\boldsymbol{\omega}:=(\lambda, \eta, a, b, c) \in \Omega \subset R^5$, $\mathbf{z}_0:=(z^1_0, \cdots, z^{10}_0) \in \mathcal{Z} \subset R^{10}$.

\section{Model calibration with deep learning}
\label{section:dl}
The Markovian nature of Model (\ref{eq:V_n}-\ref{eq:Z_factor}) makes the calibration with Monte-Carlo simulations more feasible.
However, besides parameters $\boldsymbol{\omega}$, the initial state of factors $\mathbf{z}_0$ is also supposed to be calibrated
from market data. In this case pricing with Monte-Carlo is not well adapted to classical optimal parameter search methods for model
calibration, as it leads to heavy computation and the results are not always satisfactory.
To bypass this ``curse of dimensionality", we apply deep learning to speed up further model calibration.
Deep learning has already achieved remarkable success with high-dimensional data like images and audio. 
Recently its potentials for model calibration in quantitative finance
has been investigated for example in \cite{bayer2019deep, hernandez2016model, horvath2021deep}. Two types of methods are proposed in the literature:
\begin{itemize}
  \item From prices to model parameters (PtM) \cite{hernandez2016model}: deep neural networks are trained to approximate the mapping
   from prices of some financial contracts, \textit{e.g.} options, to model parameters. With this method, we can get directly the 
   calibrated parameters from market data, without use of some numerical methods searching for optimal parameters. 
  \item From model parameters to prices (MtP) \cite{bayer2019deep, horvath2021deep}: in the first step, deep neural networks are 
  trained to approximate the pricing function, that is the mapping from model parameters to prices of financial contracts. 
  Then in the second step, traditional optimization algorithms can be used to find optimal parameters, to minimize the 
  discrepancy between market data and model outputs.
\end{itemize}
It is hard to say that one method is always better than the other. PtM method is faster and avoids computational 
errors caused by optimization algorithms, while MtP method is more robust to the varying nature of options data (strike, maturity, $\ldots$). 
In the following, the two methods are applied with implied volatility surfaces (IVS), represented by
certain points with respect to some predetermined strikes and maturities. During model calibration, all these points need to be built
from market quotes for PtM method, while we could 
focus on some points of interest for MtP method, for example those near-the-money. 
For comparison, we will test both methods in the following with simulated data. 
\subsection{Methodology}
\label{sc:simulation}
The neural networks used in our tests are all multilayer perceptrons.  
They are trained with synthetic dataset generated from the model. Our methodology is mainly based on two steps: 
\textbf{data generation} and \textbf{model training}.

\vskip 0.2in
\noindent \textbf{- Synthetic data generation}
\vskip 0.15in
\noindent The objective is to generate data samples $\big\{\boldsymbol{\omega}\footnotemark,
\footnotetext{In our tests we do not calibrate $\alpha$ and fix it to be $0.51$. In fact $c^n_i, \gamma^n_i, i=1,\cdots, n$ in $K^n$ only depend on $\alpha$, 
so making $\alpha$ constant fixes also $c^n_i, \gamma^n_i, i=1,\cdots, n$. Through experiments with market data, we find $\alpha=0.51$ is a consistently relevant choice.
Results in this paper are not sensitive to the choice of $\alpha$, and in practice we can generate few samples with other $\alpha$ and train neural networks with 
``transfer learning''. One example is given in Appendix \ref{ap:transfer}.}
\mathbf{z}_0\footnotemark, 
\footnotetext[3]{We do not include $S_0$ since we look at prices of options with respect to log-moneyness strikes.}
IVS_{SPX}, IVS_{VIX}\big\}$,
where $IVS_{SPX}$ and $IVS_{VIX}$ stand for implied volatility surface of SPX
and VIX options respectively. We randomly sample $\boldsymbol{\omega}$ and $\mathbf{z}_0$ with the following distribution:
\begin{align*}
   \boldsymbol{\omega} &\in  \mathcal{U}[0.5, 2.5]\times \mathcal{U}[1.0, 1.5]\times\mathcal{U}[0.1,0.6]\times\mathcal{U}[0.01, 0.5]\times\mathcal{U}[0.0001, 0.03] \, ,\\
   \mathbf{z}_0 &\in   \mathcal{U}[-0.5, 0.5]^{10} \, , \text{where } \mathcal{U} \text{ denotes the uniform distribution} \, .
\end{align*}
\noindent For each sampling from above distribution, we generate 50,000 random paths of SPX and VIX with the explicit-implicit 
Euler scheme (\ref{eq:euler_scheme}). Then Monte-Carlo prices are used to compute $IVS_{SPX}^{MC}$ and $IVS_{VIX}^{MC}$ with 
respect to some predetermined log-moneyness strikes and maturities:
\begin{itemize}
  \item log-moneyness strikes of SPX options: $k_{SPX}$ = \{-0.15, -0.12, -0.1, -0.08, -0.05, -0.04, -0.03, -0.02, -0.01, 0.0, 0.01, 0.02, 0.03, 0.04, 0.05\},
  \item log-moneyness strikes of VIX options: $k_{VIX}$ = \{-0.1, -0.05, -0.03, -0.01, 0.01, 0.03, 0.05, 0.07, 0.09, 0.11, 0.13, 0.15, 0.17, 0.19, 0.21\},
  \item maturities $T$ = \{0.03, 0.05, 0.07, 0.09\}.
\end{itemize}
Then $IVS_{SPX}^{MC}$ is represented by a vector of size $m=\#k_{SPX}\times\# T$, where $\# A$ is the cardinality of set $A$. 
We use flattened vectors instead of matrices as the former is more adapted to multilayer perceptrons, and analogously
for $IVS_{VIX}^{MC}$. We generate in total 150,000 data pairs as training set, 20,000 data pairs as validation set,
which is used for early stopping to avoid overfitting neural networks, and 10,000 pairs as test set for evaluating 
the performance of trained neural networks. 

\vskip 0.2in
\noindent \textbf{- Model training}
\vskip 0.15in

\begin{figure}[!h]
  \centering
  \begin{subfigure}{.5\textwidth}
      \centering
      \includegraphics[width=\linewidth]{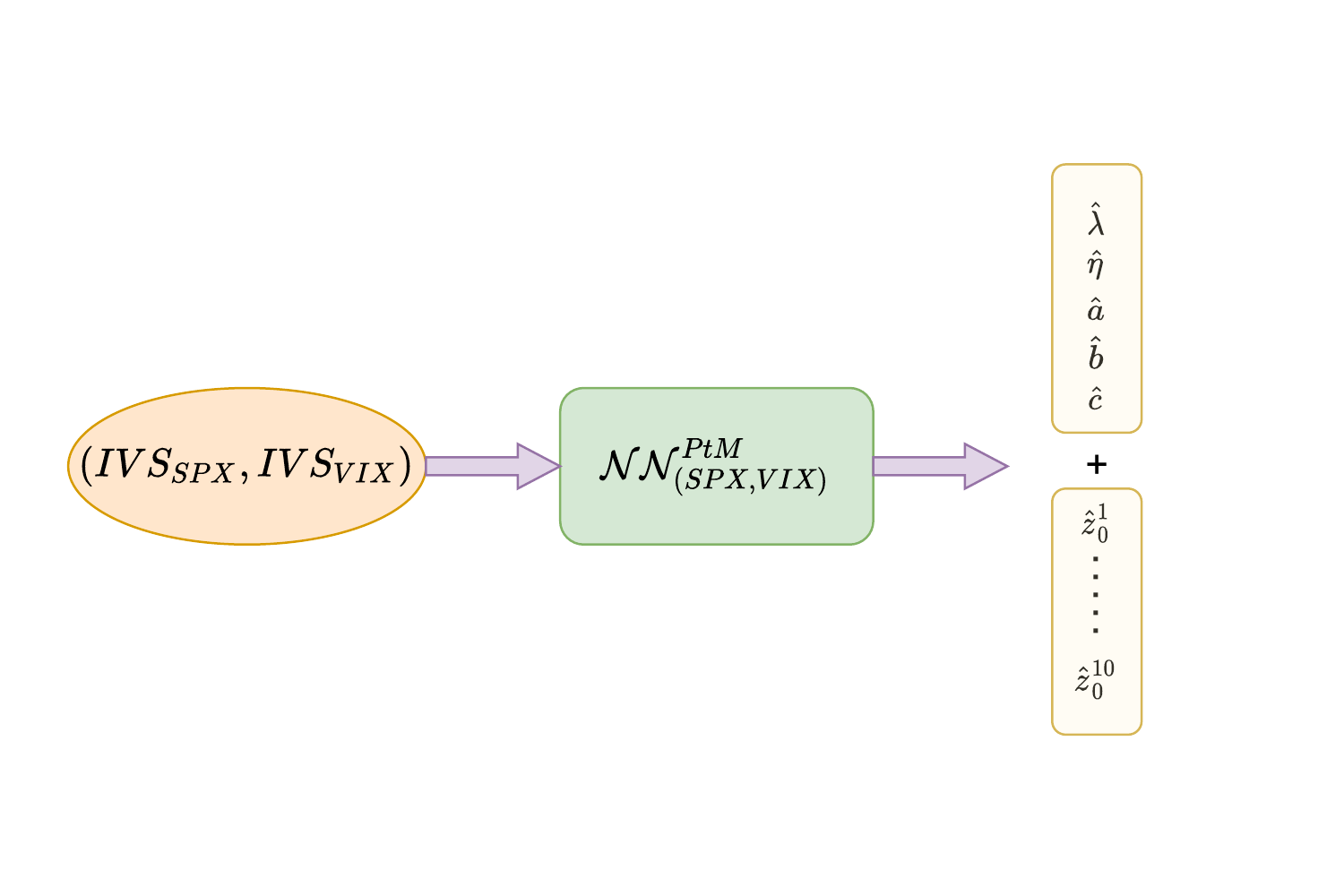}
  \end{subfigure}%
  \begin{subfigure}{.5\textwidth}
    \centering
    \includegraphics[width=\linewidth]{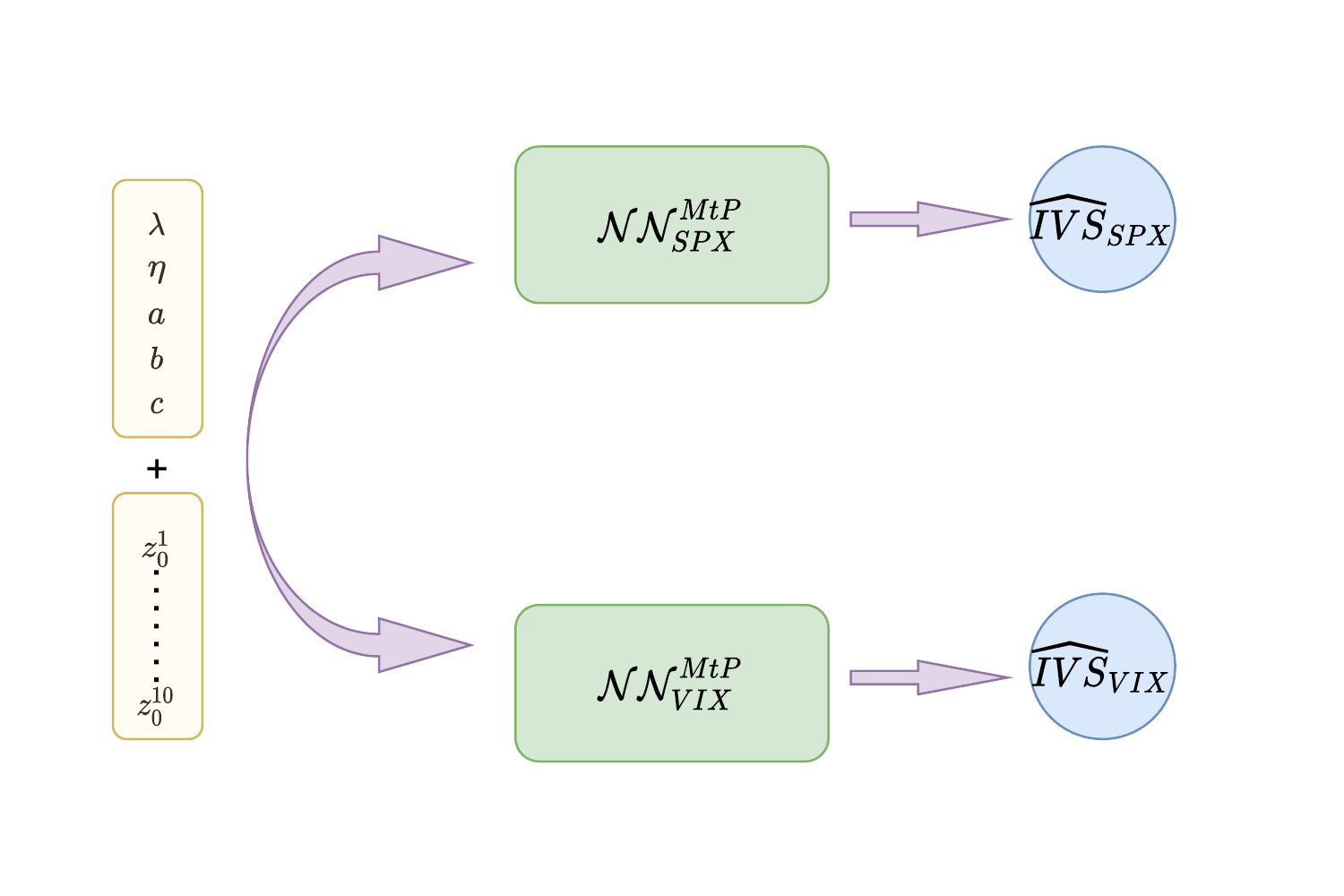}
  \end{subfigure}
  \caption{Scheme of PtM and MtP method.}
  \label{fig:schema_nn}
\end{figure}
\noindent We denote the neural network of PtM method by $\mathcal{NN}^{PtM}_{(SPX, VIX)}:\text{R}_{+}^{2m}\mapsto \Omega\times\mathcal{Z}$.
It takes $IVS_{SPX}$ and $IVS_{VIX}$ as input, and outputs estimation of parameters $\boldsymbol{\omega}$ and $\mathbf{z}_0$. 
As for MtP method, the network consists of two sub-networks, denoted with $\mathcal{NN}^{MtP}_{SPX}:\Omega\times\mathcal{Z}\mapsto\text{R}_{+}^m$ 
and $\mathcal{NN}^{MtP}_{VIX}:\Omega\times\mathcal{Z}\mapsto\text{R}_{+}^m$. They aim at approximating
the mappings from model parameters to $IVS_{SPX}$ and $IVS_{VIX}$ respectively. The methodology is illustrated in Figure \ref{fig:schema_nn}. 
Table \ref{tab:nn_summary} summarizes some key characteristics of these networks and the training process\footnote{
  We actually do not have the same number of parameters for the neural networks of the two methods. In fact it is found 
  empirically that the depth of network plays a more important role than the number of parameters, see \cite{goodfellow2016deep}. Besides, results
  presented here are not sensitive to the width of hidden layers. 
}. 
Note that for model training, each element of $\boldsymbol{\omega}$ and $\mathbf{z}_0$ is standardized to be in $[-1, 1]$, 
every point of the $IVS^{MC}$ is subtracted by the sample mean, and divided by the sample standard deviation across training set. 
\begin{table}[!h]
  \centering
  \begin{tabular}{|C{3cm}|C{3cm}|C{3cm}|C{3cm}|}
    \hline
       & $\mathcal{NN}^{PtM}_{(SPX, VIX)}$ & $\mathcal{NN}^{MtP}_{SPX}$ & $\mathcal{NN}^{MtP}_{VIX}$\\
    \hline
      Input dimension & 120 & 15 & 15 \\
    \hline
      Output dimension & 15 & 60  & 60  \\
    \hline
      Hidden layers & \multicolumn{3}{C{9cm}|}{7 with 25 hidden nodes for each, followed by SiLU activation function, 
      see \cite{hendrycks2016gaussian}} \\
    \hline
    Training epochs & \multicolumn{3}{C{9cm}|}{150 epochs with early stopping if not improved on validation set for 5 epochs} \\
    \hline
    Others & \multicolumn{3}{C{9cm}|}{Adam optimizer, initial learning rate 0.001, reduced by a factor of 2 every 10 epochs, mini-batch size 128} \\
    \hline 
  \end{tabular}
  \caption{Some key characteristics of the networks and the training process.}
  \label{tab:nn_summary}
\end{table}


\subsection{Pricing}
\begin{figure}[!h]
  \centering
  \includegraphics[width=\textwidth]{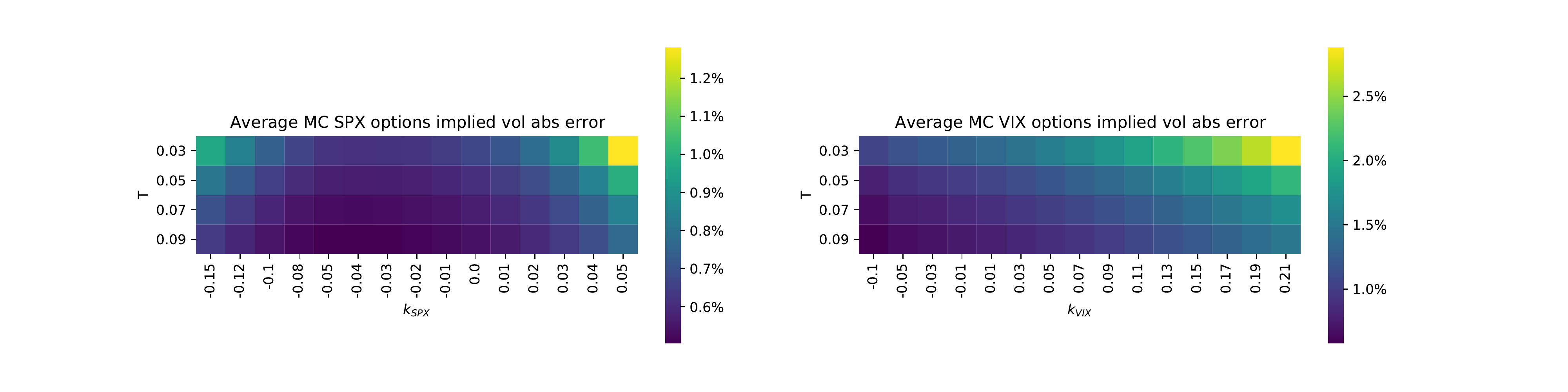}
  \caption{Average Monte-Carlo absolute errors of implied volatilities across test set, defined as the half 95\% confidence interval of Monte-Carlo simulations.}
  \label{fig:mc_spx_vix_err}
\end{figure}
\noindent In this part we first check the ability of neural networks to approximate the pricing function of the model, \textit{i.e.} the mapping from
model parameters to $IVS_{SPX}$ and $IVS_{VIX}$. To see this, we compare the estimations $\widehat{IVS}_{SPX}$ and $\widehat{IVS}_{VIX}$, given by 
$\mathcal{NN}_{SPX}^{MtP}$ and $\mathcal{NN}_{VIX}^{MtP}$, with the ``true" counterparts given by Monte-Carlo method. 

\begin{figure}[!h]
  \centering
  \includegraphics[width=\textwidth]{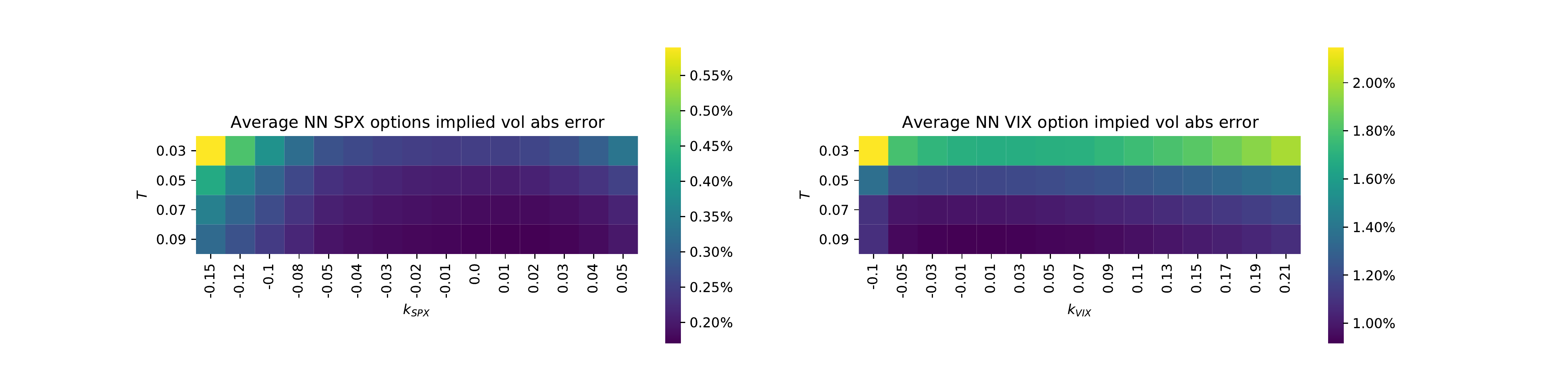}
  \caption{Average absolute errors of implied volatilities across test set for $\mathcal{NN}^{MtP}_{SPX}$ and $\mathcal{NN}^{MtP}_{VIX}$. 
      Errors are calculated with respect to Monte-Carlo counterparts.}
  \label{fig:nn_spx_vix_err}
\end{figure}

\vskip 0.2in
\noindent
Figure \ref{fig:mc_spx_vix_err} presents the benchmark, given by the half 95\% confidence interval of Monte-Carlo simulations for each points
on $IVS_{SPX}$ and $IVS_{VIX}$.
We then apply $\mathcal{NN}^{MtP}_{SPX}$ and $\mathcal{NN}^{MtP}_{VIX}$ on test set and evaluate the average absolute errors $\big(\text{abs}(\widehat{IVS}_I - IVS_I^{MC})\big)_{I=SPX, VIX}$.
The results are shown in Figure \ref{fig:nn_spx_vix_err}.
We can see that the estimations given by neural networks are close to Monte-Carlo references, 
with the majority of points of IVS falling in the 95\% confidence interval of Monte-Carlo. We also tested average relative errors as an alternative metric,
defined as $\big(\text{abs}(\widehat{IVS}_I - IVS_I^{MC})\oslash IVS_I^{MC}\big)_{I=SPX, VIX}$, where $\oslash$ means element-wise division between vectors.
The results are given in Figure \ref{fig:mc_spx_vix_rel_err} and Figure \ref{fig:nn_spx_vix_rel_err} in Appendix,
and are consistent with the observations in Figures \ref{fig:mc_spx_vix_err}-\ref{fig:nn_spx_vix_err}.

\vskip 0.2in
\noindent
At this stage, it is reasonable to conclude that $\mathcal{NN}^{MtP}_{SPX}$ and $\mathcal{NN}^{MtP}_{VIX}$ are able to learn the pricing functions from data of Monte-Carlo simulations.
With these networks we can generate implied volatility surfaces for SPX and VIX for any model parameters.

\subsection{Calibration}
\begin{figure}[!h]
  \centering
  \includegraphics[width=\textwidth]{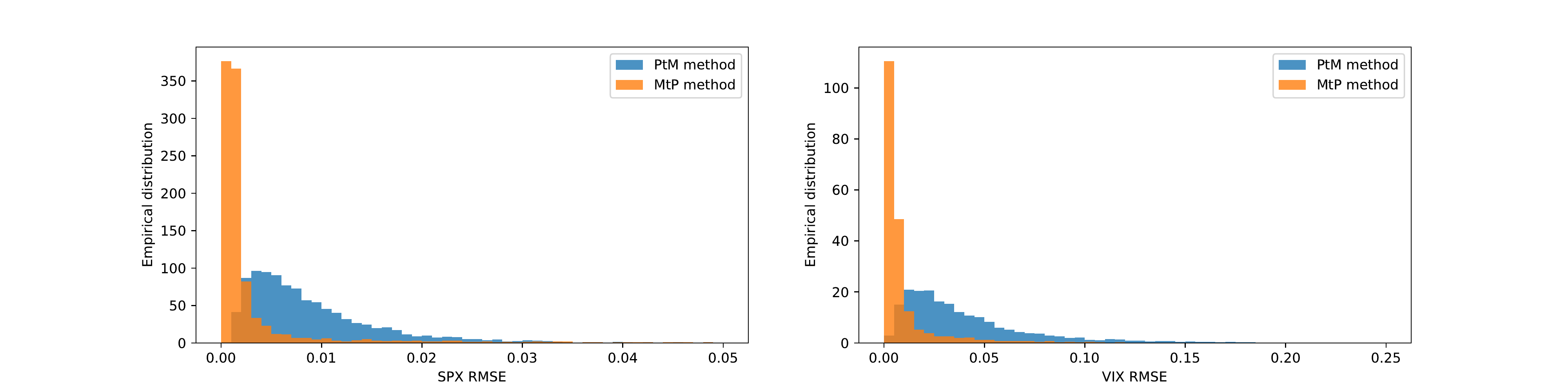}
  \caption{Empirical distribution of reconstruction RMSE of IVS.}
  \label{fig:nn_calib_rmse}
\end{figure}

In this part, we use $\mathcal{NN}^{PtM}_{(SPX, VIX)}$, $\mathcal{NN}^{MtP}_{SPX}$ and $\mathcal{NN}^{MtP}_{VIX}$ to perform model calibration.
For PtM method, the output of network gives directly calibration results. For MtP method, the result is given as 
\begin{equation}
  \hat{\boldsymbol{\omega}}, \hat{\mathbf{z}}_0 = \text{argmin}_{\boldsymbol{\omega}\in \Omega,\mathbf{z}_0\in\mathcal{Z}} \, \|\mathcal{NN}^{MtP}_{SPX}(\boldsymbol{\omega},\mathbf{z}_0) - IVS_{SPX}\|^2_2 + \|\mathcal{NN}^{MtP}_{VIX}(\boldsymbol{\omega},\mathbf{z}_0) - IVS_{VIX}\|^
  2_2 \, .
\label{eq:calib_obj}
\end{equation}
where $\Omega=[0.5, 2.5]\times[1.0, 1.5]\times[0.1, 0.6]\times[0.01, 0.5]\times[0.0001, 0.03]$, $\mathcal{Z}=[-0.5, 0.5]^{10}$. 
We use the same weight for all points of IVS in (\ref{eq:calib_obj}). In practice we could consider different weights to adapt 
to the varying nature of market quotes in terms of liquidity, bid-ask spread, etc. 
We apply L-BFGS-B algorithm for this optimization problem. Note that it is a gradient-based algorithm, and the gradients needed are
calculated directly with $\mathcal{NN}_{SPX}^{MtP}$ and $\mathcal{NN}_{VIX}^{MtP}$ via \textit{automatic adjoint differentiation} (AAD), which is 
already implemented in popular deep learning frameworks like TensorFlow and PyTorch. One can refer to Section \ref{sc:delta_calc} for calculation principles.
\vskip 0.2in
\noindent
\textbf{Calibration on simulated data}
\vskip 0.15in
\noindent
We apply the two methods on test set generated by Monte-Carlo simulations, and
we evaluate the calibration by Normalized Absolute Errors (NAE) of parameters and the reconstruction
Root Mean Square Errors (RMSE) of IVS:
\begin{align*}
  \text{NAE} &= \frac{|\hat{\theta} - \theta|}{\theta_{up} - \theta_{low}} \, , \\
  \text{RMSE} &= \sqrt{\frac{1}{\#k_I\times \#T}\|\mathcal{NN}^{MtP}_I(\hat{\boldsymbol{\omega}}, \mathbf{\hat{z}}_0) - IVS^{MC}_I\|_2^2}, \quad I \in \{\text{SPX}, \text{VIX}\} \, ,
\end{align*}
where $\theta$ is one element of $\boldsymbol{\omega}$ or $\mathbf{z}_0$, and $\theta_{up}, \theta_{low}$ stand for the upper bound and lower bound of the uniform distribution for sampling $\theta$. 
Note that we could alternatively use Monte-Carlo to reconstruct IVS with the calibrated parameters instead of $\mathcal{NN}^{MtP}_{SPX}$ and 
$\mathcal{NN}^{MtP}_{VIX}$. However it would be much slower and we have seen in the above
that the outputs given by $\mathcal{NN}^{MtP}_{SPX}$ and $\mathcal{NN}^{MtP}_{VIX}$ are very close to those of Monte-Carlo. 
Figure \ref{fig:nn_calib_err} in Appendix shows the empirical cumulative distribution function of NAE for all the calibrated parameters.
Figure \ref{fig:nn_calib_rmse} gives the empirical distribution of RMSE of IVS with the calibrated parameters from the two methods.
\noindent We can make the following remarks:
\begin{itemize}
  \item From Figure \ref{fig:nn_calib_err} in Appendix, we see that PtM method can usually get smaller discrepancy for calibrated parameters
        than MtP method. This is not surprising since the latter may end with locally optimal solutions when we use gradient-based optimization algorithms.
  \item MtP method performs better in terms of reconstruction error, which is expected since it is exactly what the algorithm tries to minimize.
  \item Our results are of course not as accurate as those reported in \cite{horvath2021deep} for some other rough volatility models, 
        especially in terms of calibration errors for model parameters.
        This is because our model is more complex and has more parameters. Consequently more data and more complicated network architecture are demanded for training. 
        The algorithms are also more likely to end with locally optimal solutions.  
\end{itemize}

\begin{figure}[!h]
  \centering
  \includegraphics[width=0.8\textwidth]{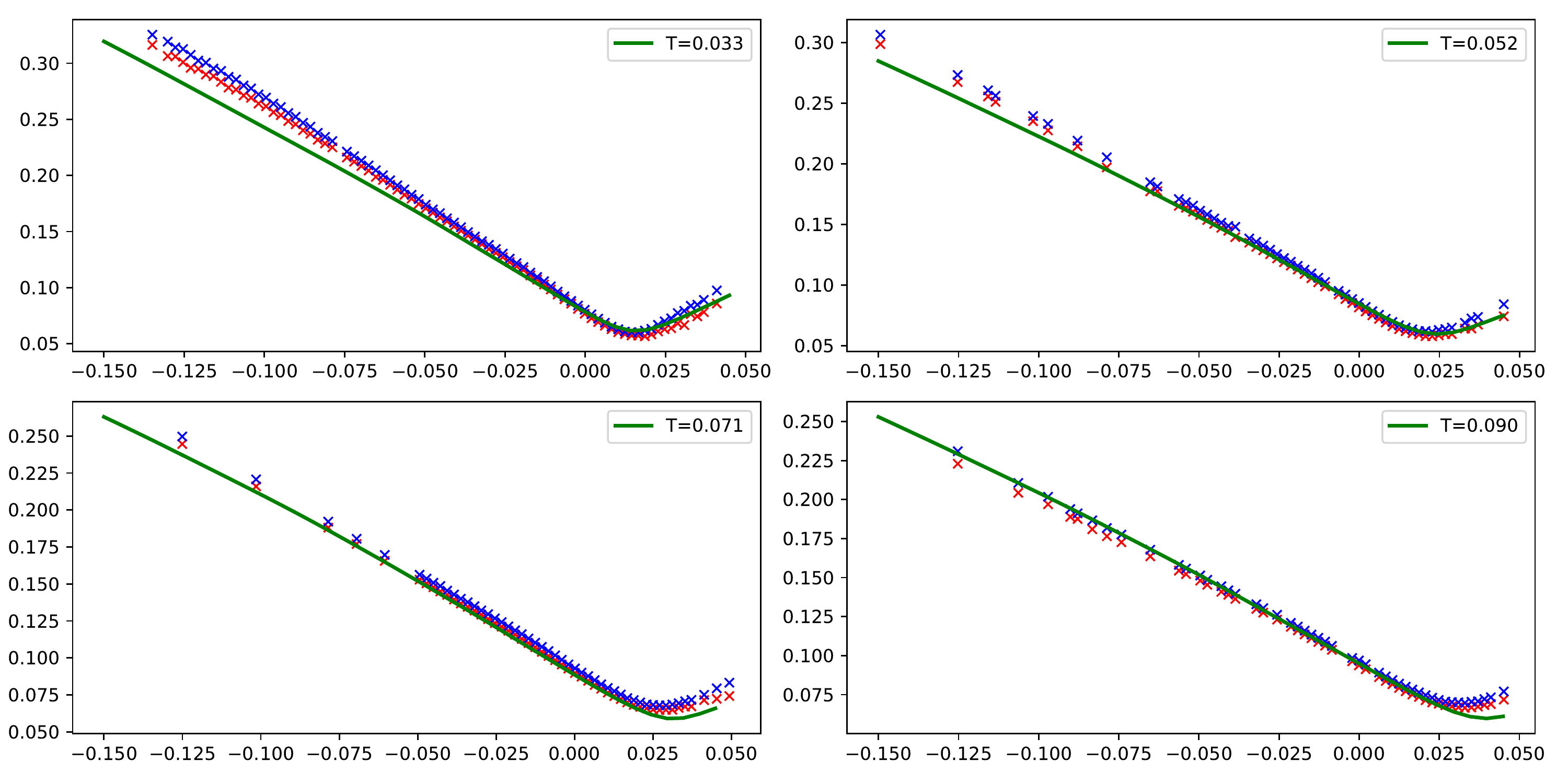}
  \caption{Implied volatilities on SPX options for 19 May 2017. Bid and ask of market volatilities are represented respectively 
        by red and blue points. Green line is the output of model with Monte-Carlo method.}
  \label{fig:spx_imp_vol_sample}
\end{figure}

\begin{figure}[!h]
  \centering
  \includegraphics[width=0.8\textwidth]{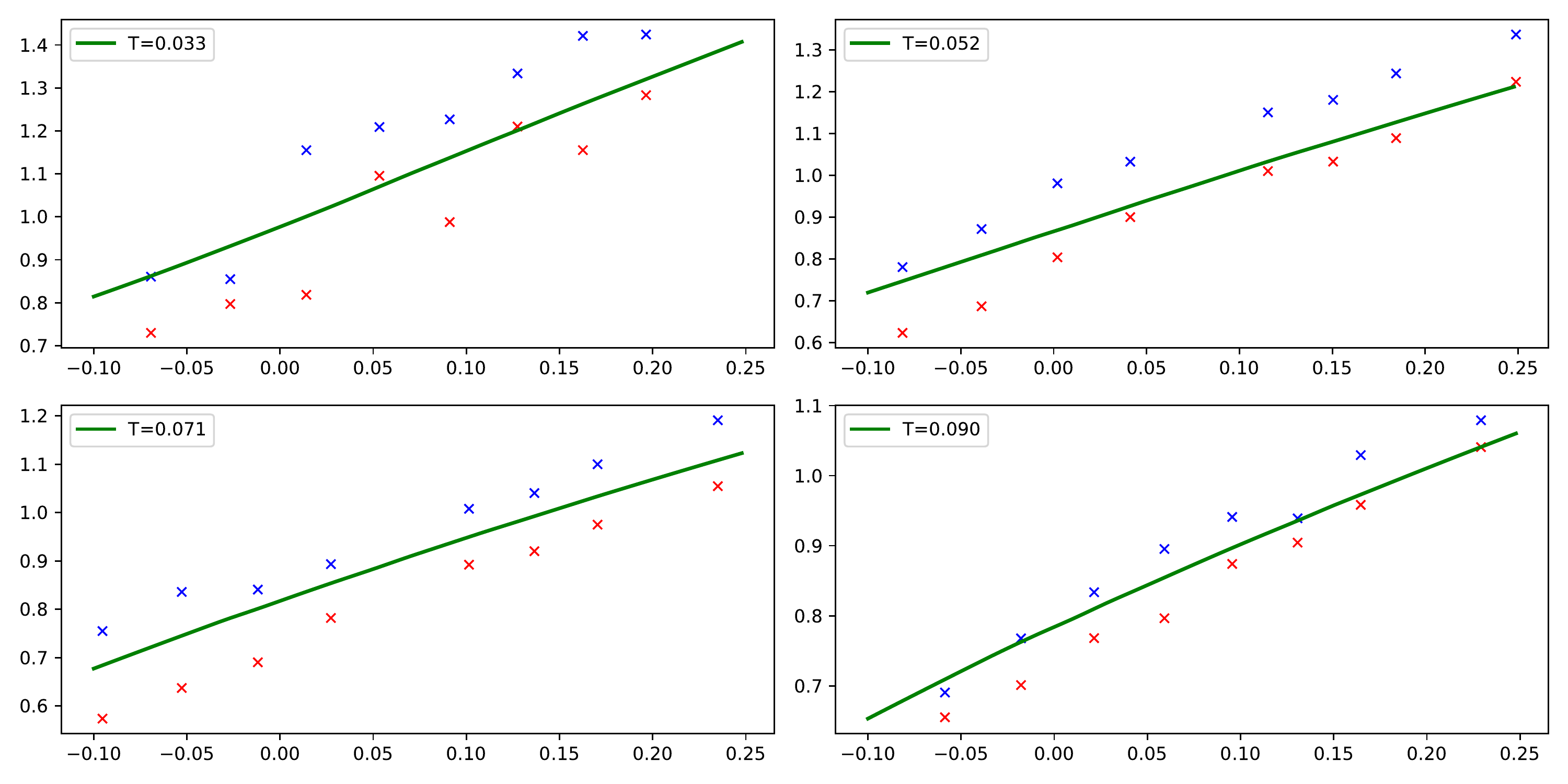}
  \caption{Implied volatilities on VIX options for 19 May 2017. Bid and ask of market volatilities are represented respectively 
        by red and blue points. Green line is the output of model with Monte-Carlo method.}
  \label{fig:vix_imp_vol_sample}
\end{figure}

\vskip 0.2in
\noindent
\textbf{Calibration on market data}
\vskip 0.15in
\noindent
We use MtP approach on market data since there are no ``true" reference model parameter values in this case, and the objective is to minimize the
discrepancy between model outputs and market observations. The arbitrage-free implied volatility interpolation method presented 
in \cite{gatheral2014arbitrage} is used to generate IVS with the same log-moneyness strikes and maturities as before. 
Taking the data of 19 May 2017 tested in \cite{gatheral2020quadratic} as example, we get the following calibration results:
\begin{align*}
   \boldsymbol{\omega} &= (2.5, 1.485, 0.401, 0.235, 0.001), \\ 
   \mathbf{z}_0 &=(-0.033, 0.015, -0.004, 0.017, 0.028, 0.098, 0.192, -0.076, 0.072, -0.062) \, .
\end{align*}
We then use Monte-Carlo to get whole IVS of SPX and VIX with these parameters. The slices corresponding to several existing maturities in the market
are shown in Figure \ref{fig:spx_imp_vol_sample} and Figure \ref{fig:vix_imp_vol_sample}.

\vskip 0.2in
\noindent
Our model fits very well the globe shape of IVS of SPX and VIX options at the same time. Model's outputs fall essentially
between bid and ask quotes for VIX options. In addition, excellent fits are obtained in terms of at-the-money skew of SPX options
\footnote{For quotes far from the money, we can remark a discrepancy between the market and the output with Monte-Carlo method. 
In fact, the mismatch between the interpolated implied volatility surface and the market on these points, and the deviation of Monte-Carlo means 
from neural networks' outputs can both induce this discrepancy.}. 
Note that we do not require the market quotes of SPX options and VIX options to have
same maturities. Another example of joint calibration is given in Appendix \ref{app:calib_sample_1}.

\vskip 0.2in
\noindent Figure \ref{fig:market_calib_params} in Appendix presents the historical dynamics of parameters from daily calibration 
on market data\footnote{In this paper we limited the parameters in some restricted intervals to illustrate the methodology with 
reasonable size of random sampling. The observation that the predetermined bounds are reached for some parameters
indicates that these intervals cannot cover all market situations. Interested readers can choose wider ones or use unbounded 
distribution like Gaussian for random sampling to relax this issue, although it may demand more synthetic data for network training.}. It is interesting to remark that during the beginning of COVID-19 crisis, $a, b, c$ all increased,
which means stronger feedback effect, more distinct feedback asymmetry and larger base variance level. The quantity 
$Z_0:=\sum_{i=1}^{10}c_iz^i_0$ became negative. All these changes contributed to larger volatility.
\begin{figure}[!h]
  \centering
  \includegraphics[width=\textwidth]{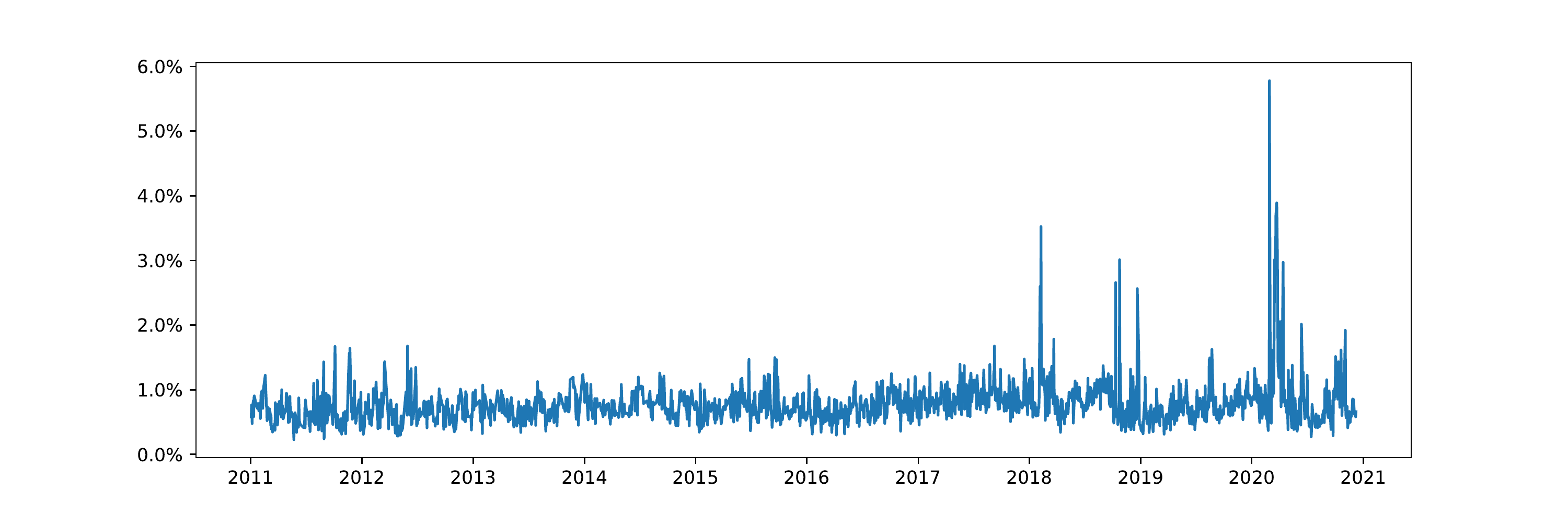}
  \caption{Historical reconstruction RMSE of $IVS_{SPX}$.}
  \label{fig:spx_ivs_rmse}
\end{figure}
\begin{figure}[!h]
  \centering
  \includegraphics[width=\textwidth]{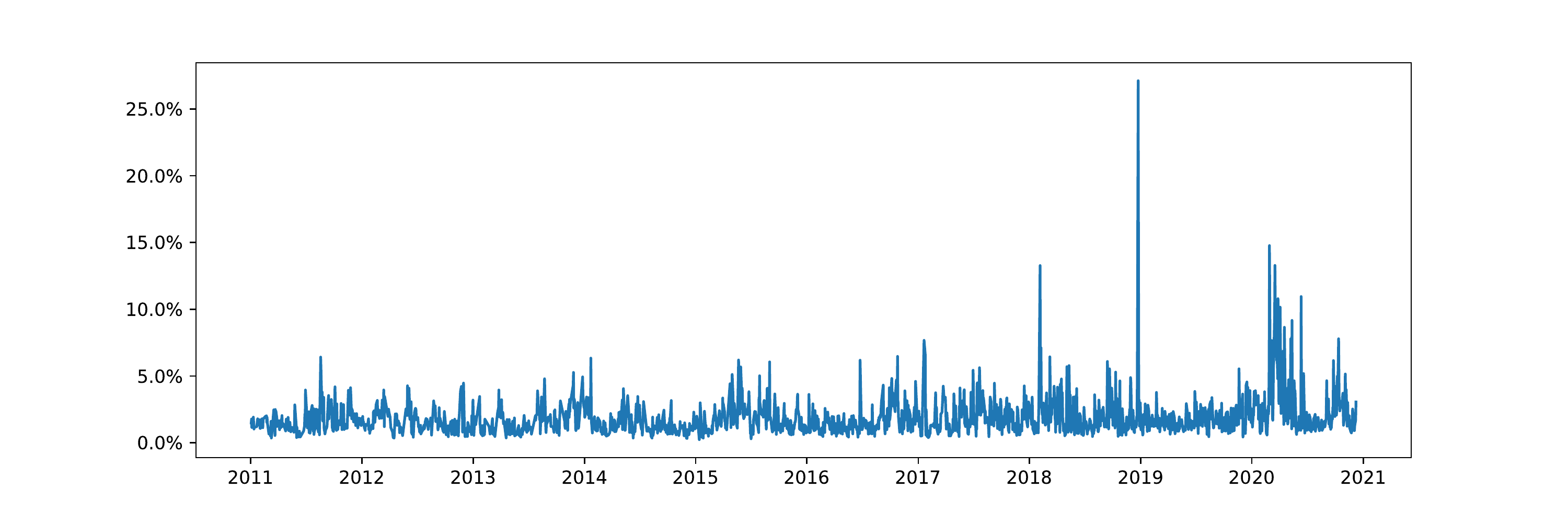}
  \caption{Historical reconstruction RMSE of $IVS_{VIX}$.}
  \label{fig:vix_ivs_rmse}  
\end{figure}

\vskip 0.2in
\noindent
Figures \ref{fig:spx_ivs_rmse} and \ref{fig:vix_ivs_rmse} show historical RMSE for $IVS_{SPX}$ and $IVS_{VIX}$. 
For most time periods, the model can fit very well the two IVS at the same time, with small RMSE. For certain dates with
extremely large market moves, we observe some spikes of errors. In fact for these dates market liquidity is usually very concentrated on
one type of contract (Call or Put) with specific strikes and maturities. In this case, market IVS is not smooth enough so that
the model may fail to output satisfying fit. Alternative more robust IVS interpolation methods could be tested in practice to generate smoother IVS. We
could also focus only on some points of interest on IVS by excluding other points from the calibration objective (\ref{eq:calib_obj}).
Note that all the empirical results presented here are with $\alpha=0.51$ and $n=10$. Interested readers can try other $\alpha$ or bigger $n$ to
improve globally historical RMSE, while new synthetic data need to be generated for each new setting.

\section{Toy examples for hedging}
\label{section:hedging}
We have seen in the previous section that the proposed model can jointly fit $IVS_{SPX}$ and $IVS_{VIX}$ with small errors. 
Then it is important to know how to hedge options with this model. In this section, we give toy examples on synthetic data and 
market data as well to show how to use the neural networks to perform hedging for vanilla SPX calls. 
We will see that in our model perfect hedging for these products is possible with only SPX. 

\subsection{Hedging portfolio computation with neural networks}
\label{sc:delta_calc}
Let $\mathbf{Z}_t:=(Z^1_t,\cdots,Z^{10}_t)$ and $\mathbf{X}_t:=(S_t, \mathbf{Z}_t)$. As indicated in Remark (\ref{re:markov_model}), Model (\ref{eq:V_n}-\ref{eq:Z_factor})
being Markovian, given strike $K$, maturity $T$ and model parameters $\boldsymbol{\omega}$, the price of vanilla SPX call at time $t$ is then a function of $\mathbf{X}_t$.
Let $P_t(\mathbf{X}_t; K, T, \boldsymbol{\omega})$ denote this quantity. With the dynamics of $(Z^i_t)_{i=1,\cdots,10}$ in Equation (\ref{eq:Z_factor}), we then have
\begin{equation}
  dP_t(\mathbf{X}_t; K, T, \boldsymbol{\omega}) = \delta_tdS_t \, ,
\end{equation}
where
\begin{equation}
  \delta_t = \frac{\partial P_t(\mathbf{X}_t; K, T, \boldsymbol{\omega})}{\partial S_t} + \frac{\eta}{S_t}\sum_{i=1}^{10}\frac{\partial P_t(\mathbf{X}_t; K, T, \boldsymbol{\omega})}{\partial Z^i_t} \, .
  \label{eq:hedge_ratio}
\end{equation}
Note that the factors $(Z^{i})_{i=1,\cdots,10}$ can be fully traced because they are assumed to be driven by the same Brownian motion as $S$, which is observable from market data.
With the neural networks approximating the pricing function of the model, we will see that we can then obtain approximation of $\delta_t$ for any $t$.
Of course continuous hedging is impossible in practice. Here we perform discrete hedging with time step $\Delta t$.
The Profit and Loss (P\&L) of hedging at $t\in(0,T]$ is given by
\begin{equation}
  \mathcal{J}_t =\mathcal{J}^{\delta}_t - \mathcal{J}^{P}_t,  
  \label{eq:hedge_pay}
\end{equation}
where
\begin{align*}
  \mathcal{J}^{\delta}_t &= \sum_{k=0}^{\floor*{t/\Delta t}-1}\hat{\delta}_{t_k} \triangle S_{t_{k+1}} + \hat{\delta}_{t_{\floor*{t/\Delta t}}}(S_t - S_{\floor*{t/\Delta t}\Delta t}) \, ,\\
  \mathcal{J}^P_t &= P_t - P_0\, ,
\end{align*}
with $t_k = k\triangle t$, $\triangle S_{t_k}:= S_{t_k} - S_{t_{k-1}}$, $\hat{\delta}_{t_k}$ is the hedging ratio given by neural networks at $k$-th hedging time $t_k$,
and $P_t$ is the price of the SPX option to hedge at time $t$. As we can see, $\mathcal{J}^\delta$ stands for the P\&L coming from holding the underlying
and $\mathcal{J}^{P}$ reflects the price evolution of the option.
We show in the following that $\hat{\delta}_{t_k}$ can be given directly from $\mathcal{NN}_{SPX}^{MtP}$ in our model.

\vskip 0.2in
\noindent
Note that $\mathcal{NN}_{SPX}^{MtP}$ behaves like a ``global'' pricer that is reusable under any model parameters $\boldsymbol{\omega}$. Given $\boldsymbol{\omega}$,
we could actually train a finer network as a ``local'' pricer taking only $\mathbf{z}_0$ as input. Of course the same methodology as before
can be used with fixed $\boldsymbol{\omega}$. Here we apply alternatively Differential Machine Learning as a fast method to obtain approximation
of pricing function from simulated paths, under a given calibration of the model, see \cite{huge2020differential}.
\vskip 0.2in
\noindent
\textbf{Method 1: with} $\mathcal{NN}_{SPX}^{MtP}$
\vskip 0.15in
\noindent
With $\mathcal{NN}_{SPX}^{MtP}$ outputing implied volatilities with respect to log-moneyness strikes, we have
$$
  P_t(\mathbf{X}_t; K, T, \boldsymbol{\omega}) \simeq P^{BS}\big(S_t, K, T-t, \sigma_{\mathcal{NN}}^{\log(K/S_t), T-t}(\boldsymbol{\omega}, \mathbf{Z}_t)\big) \, ,
$$
where $P^{BS}(S, K, T, \sigma)$ is the price of European call under Black-Scholes model and $\sigma_{\mathcal{NN}}^{k,T}(\boldsymbol{\omega}, \mathbf{Z}_t)$ is the implied volatility
corresponding to log-moneyness strike $k$ and maturity $T$, calculated directly by $\mathcal{NN}^{MtP}_{SPX}$ with $(\boldsymbol{\omega}, \mathbf{Z}_t)$ as input.
Then the partial derivatives in (\ref{eq:hedge_ratio}) are given by
\begin{align}
  \begin{split}
    \frac{\partial P_t(\mathbf{X}_t; K, T, \boldsymbol{\omega})}{\partial S_t} &\simeq \delta_{BS}\big(S_t, K, T-t, \sigma_{\mathcal{NN}}^{\log(K/S_t), T-t}(\boldsymbol{\omega}, \mathbf{Z}_t)\big)  +  \\
                             &\nu_{BS}\big(S_t, K, T-t, \sigma_{\mathcal{NN}}^{\log(K/S_t), T-t}(\boldsymbol{\omega}, \mathbf{Z}_t)\big)\frac{\partial\sigma_{\mathcal{NN}}^{k, T-t}(\boldsymbol{\omega}, \mathbf{Z}_t)}{\partial k}\lvert_{k=\log(K/S_t)}\frac{\partial \log(K/S_t)}{\partial S_t}, \, \label{eq:nn_hedge_S}
  \end{split}
\end{align}
\begin{equation}
  \frac{\partial P_t(\mathbf{X}_t; K, T, \boldsymbol{\omega})}{\partial Z_t^{i}} \simeq \nu_{BS}\big(S_t, K, T-t, \sigma_{\mathcal{NN}}^{\log(K/S_t), T-t}(\boldsymbol{\omega}, \mathbf{Z}_t)\big)\frac{\partial\sigma_{\mathcal{NN}}^{\log(K/S_t), T-t}(\boldsymbol{\omega}, \mathbf{Z}_t)}{\partial Z^{i}_t}, \, 
  \label{eq:nn_hedge_Z}
\end{equation}
where $\delta_{BS}(S, K, T, \sigma)$ and $\nu(S, K, T, \sigma)$ stand for the \textit{Delta} and \textit{Vega} respectively under Black-Scholes model.
The quantity $\frac{\partial\sigma_{\mathcal{NN}}^{k, T}}{\partial Z^{i}_t}$ corresponds actually
to the derivative of the outputs of $\mathcal{NN}^{MtP}_{SPX}$ with respect to its inputs. Thus it can be obtained instantaneously with built-in AAD.
$\frac{\partial \sigma_{\mathcal{NN}}^{k, T-t}(\boldsymbol{\omega}, \mathbf{Z}_t)}{\partial k}$ in (\ref{eq:nn_hedge_S}) can be approximated by the finite difference $\frac{(\sigma^{k+\delta_k, T-t}_{\mathcal{NN}} - \sigma^{k-\delta_k, T-t}_{\mathcal{NN}})}{2\delta_k}(\boldsymbol{\omega}, \mathbf{Z}_t)$. 
Note that some interpolation methods need to be applied for arbitrary pair $(K, T-t)$ since $\mathcal{NN}^{MtP}_{SPX}$ has fixed log-moneyness strikes
and maturities. 

\vskip 0.2in
\noindent
\textbf{Method 2: with Differential Machine Learning}
\vskip 0.15in
\noindent
Given parameters $\boldsymbol{\omega}$, we can simulate a path of model state $(\mathbf{X}_t)_{0\leq t \leq T}$ starting from the initial state $\mathbf{X}_0$. 
The pathwise payoff $(S_T - K)_+$ is in fact an unbiased estimation of $P_0(\mathbf{X}_0; K, T, \boldsymbol{\omega})$. Under some regularity conditions, 
the pathwise derivative $\frac{\partial (S_T - K)_+}{\partial \mathbf{X}_0}$ is also an unbiased estimation of $\frac{\partial P_0(\mathbf{X}_0; K,T, \boldsymbol{\omega})}{\partial \mathbf{X}_0}$.
We show in the following how to calculate this quantity with the simulation scheme proposed in (\ref{eq:euler_scheme}). The basic idea of Differential Machine Learning
\cite{giles2006smoking,huge2020differential} is to concatenate pathwise
payoff and pathwise derivatives as targets to train a neural network, denoted by $\mathcal{NN}_{DML}$, to approximate the pricing mapping from $\mathbf{X}_0$ to $P_0$ under some fixed $\boldsymbol{\omega}$. 
Thus, the training samples are like $\big\{\mathbf{X}_0, \big((S_T - K)_+, \frac{\partial(S_T-K)_+}{\partial \mathbf{X}_0}\big)\big\}$, and the loss function for training
is like
\begin{equation}
    \mathcal{L} = \mathcal{L}_1\big(\mathcal{NN}_{DML}(\mathbf{X}_0), (S_T - K)_+\big) + \mathcal{L}_2\big(\frac{\partial\mathcal{NN}_{DML}(\mathbf{X}_0)}{\partial\mathbf{X}_0}, \frac{\partial(S_T-K)_+}{\partial \mathbf{X}_0}\big) \, ,
    \label{eq:dml_loss}
\end{equation}
with $\mathcal{L}_1$ and $\mathcal{L}_2$ some suitably chosen loss functions. Similarly to the case with $\mathcal{NN}^{MtP}_{SPX}$, $\frac{\partial\mathcal{NN}_{DML}(\mathbf{X}_0)}{\partial\mathbf{X}_0}$
can be calculated efficiently with AAD.  
In this way, $\mathcal{NN}_{DML}$ aims at learning both the pricing function and its derivatives during training. This can help the networks converge with few samples, see \cite{huge2020differential}. 

\vskip 0.2in
\noindent
The quantity $\frac{\partial (S_T - K)_+}{\partial \mathbf{X}_0}$ can be calculated with AAD, which is based on the chain rule of derivatives computation, see \cite{giles2006smoking} for more details.
In our case, with the Euler scheme in (\ref{eq:euler_scheme}), let $\Delta t$ the simulation step with $N\Delta t =T$.
We have
\begin{align*}
  \frac{\partial \hat{S}_{k+1}}{\partial X^j_0} &= \big(1 + \sqrt{\hat{V}_k}(W_{k+1} - W_k)\big)\frac{\partial \hat{S}_k}{\partial X^j_0} + \frac{a\hat{S}_{k}(W_{k+1} - W_k)(\sum_{i=1}^{10}c_i\hat{Z}^{i}_k - b)}{\sqrt{\hat{V}_k}}\sum_{i=1}^{10}c_i\frac{\partial\hat{Z}^{i}_k}{\partial X^j_0}  \, , \\
  \frac{\partial\hat{Z}^{i}_{k+1}}{\partial X^j_0} &= \frac{1}{1+\gamma_i\Delta t} \big(\frac{\partial\hat{Z}^{i}_k}{\partial X^j_0} + (-\lambda\Delta t + \frac{\eta a(W_{k+1} - W_k)(\sum_{i=1}^{10}c_i\hat{Z}^{i}_k - b)}{\sqrt{\hat{V}_k}})\sum_{i=1}^{10}c_i\frac{\partial\hat{Z}^{i}_k}{\partial X^j_0}\big) \, ,
\end{align*}
where $X^j_0$ is the $j$-th element of $\mathbf{X}_0$. Let $\mathbf{\hat{X}}_k := (\hat{S}_k, \hat{Z}^{1}_k, \cdots,  \hat{Z}^{10}_k)$.
This can be rewritten in matrix form $\boldsymbol{\Delta}(k+1) = \mathbf{D}(k)\mathbf{\Delta}(k)$, with $\mathbf{\Delta}_{i,j}(k)=\frac{\partial\hat{X}^i_k}{\partial X^j_0}, i,j=1, \cdots, 11$, and
$$
    \mathbf{D}(k) = \begin{pmatrix}
      1+\sqrt{\hat{V}_k}\Delta W_{k+1}, & \hat{S}_kM^1_kc_1, & \cdots & \hat{S}_kM^1_kc_{10}   \\ 
      0, & \frac{1+c_1M^2_k}{1+\gamma_1\Delta t}, &  \cdots & \frac{c_{10}M^2_k}{1+\gamma_{10}\Delta t} \\
      \vdots & \vdots & \vdots & \vdots \\
      0 &  \frac{c_{1}M_k^2}{1+\gamma_1\Delta t}, & \cdots & \frac{1+c_{10}M_k^2}{1+\gamma_{10}\Delta t}  
    \end{pmatrix} \, ,
$$
where 
\begin{align*}
    \Delta W_{k+1} &:= W_{k+1} - W_k \, ,\\ 
    M_k^1 &:= \frac{a\Delta W_{k+1}(\sum_{i=1}^{10}c_i\hat{Z}_k^i-b)}{\sqrt{\hat{V}_k}} \, , \\
    M_k^2 &:= \eta M^1_k - \lambda\Delta t \, . 
\end{align*}
Then we have
\begin{align}
\begin{split}
\frac{\partial (\hat{S}_T - K)_+}{\partial \mathbf{X}_0} &= \mathbf{\Delta}(N)^T\frac{\partial (\hat{S}_T - K)_+}{\partial\mathbf{\hat{X}}_N} \\
          &= \mathbf{\Delta}(0)^T\mathbf{D}(0)^T\cdots \mathbf{D}(N-2)^T\mathbf{D}(N-1)^T\frac{\partial (\hat{S}_T - K)_+}{\partial \mathbf{\hat{X}}_N} \\
          &= \mathbf{\Delta}(0)^T\mathbf{V}(0) \, ,
\end{split}
\label{eq:path_deriv}
\end{align}
where $\mathbf{\Delta}(0)$ is simply the identity matrix by definition, and $\mathbf{V}(0)$ can be calculated recursively:
\begin{align}
  \begin{split}
  \mathbf{V}(k) &= \mathbf{D}(k)^T\mathbf{V}(k+1), \quad k=0,\cdots, N-1 \, ,\\
  \mathbf{V}(N) &= \frac{\partial (\hat{S}_T - K)_+}{\partial\mathbf{\hat{X}}_N} = \mathbf{1}(\hat{S}_T > K)(1, 0, \cdots, 0)^T \, .
  \end{split}
  \label{eq:deriv_recurv}
\end{align}
\noindent
Note that for each simulated path, $\big(\mathbf{D}(k)\big)_{k=0,\cdots, N-1}$ can be readily obtained, so the quantity $\frac{\partial (\hat{S}_T - K)_+}{\partial \mathbf{X}_0}$
can be efficiently computed following (\ref{eq:path_deriv}, \ref{eq:deriv_recurv}).
\vskip 0.2in
\noindent
Since we use the same trained network for hedging at any $t\in(0, T]$, to accommodate the varying time to maturity $(T-t)$, 
we consider multiple outputs corresponding to different maturities ${T_1, T_2, \cdots, T_m}$ for $\mathcal{NN}_{DML}$, with $T_1<T<T_m$. 
The quantities $(\frac{\partial(\hat{S}_{T_i} - K)_+}{\partial \mathbf{X}_0})_{i=1,\cdots,m}$ can be computed following (\ref{eq:path_deriv}, \ref{eq:deriv_recurv}).
For training, we use the average of $m$ derivatives in (\ref{eq:dml_loss}) for simplification, see \cite{huge2020differential} for more details on designs with 
multi-dimensional output. After training $\mathcal{NN}_{DML}$ with respect to strike $K$ and parameter $\boldsymbol{\omega}$, we have
$$
P_t(\mathbf{X}_t; K, T, \boldsymbol{\omega}) \simeq \mathcal{NN}_{DML}^{T-t}(\mathbf{X}_t) \, ,
$$
with $\mathcal{NN}_{DML}^{T-t}(\cdot)$ the output corresponding to maturity $(T-t)$. Thus we get
\begin{align*}
  \frac{\partial P_t(\mathbf{X}_t; K, T, \boldsymbol{\omega})}{\partial S_t} &\simeq \frac{\partial \mathcal{NN}_{DML}^{T-t}(\mathbf{X}_t)}{\partial S_t} \, , \\
  \frac{\partial P_t(\mathbf{X}_t; K, T, \boldsymbol{\omega})}{\partial Z_t^i} &\simeq \frac{\partial \mathcal{NN}_{DML}^{T-t}(\mathbf{X}_t)}{\partial Z_t^i} \, .
\end{align*}
Then the formula in (\ref{eq:hedge_ratio}) is used to obtain hedging ratio. As in the case with $\mathcal{NN}^{MtP}_{SPX}$, some interpolation methods are needed for arbitrary $(T-t)$.
In our tests, we choose the following characteristics for $\mathcal{NN}_{DML}$ and its training:
\begin{itemize}
  \item 4 hidden layers, with 20 hidden nodes for each, SiLU as activation function,
  \item input dimension is 11, output dimension is 5 corresponding to maturities $0.02, 0.04, 0.06, 0.08, 0.1$,
  \item mini-batch gradient descent with batch size 128, initial learning rate 0.001, divided by 2 every 5 epochs,
  \item sample uniformly $\mathbf{X}_0$ and simulate 50,000 paths, train $\mathcal{NN}_{DML}$ with 20 epochs.
\end{itemize}

\subsection{Numerical results}

\begin{figure}[!h]
  \centering
  \begin{subfigure}{.5\textwidth}
    \centering
    \includegraphics[width=\linewidth]{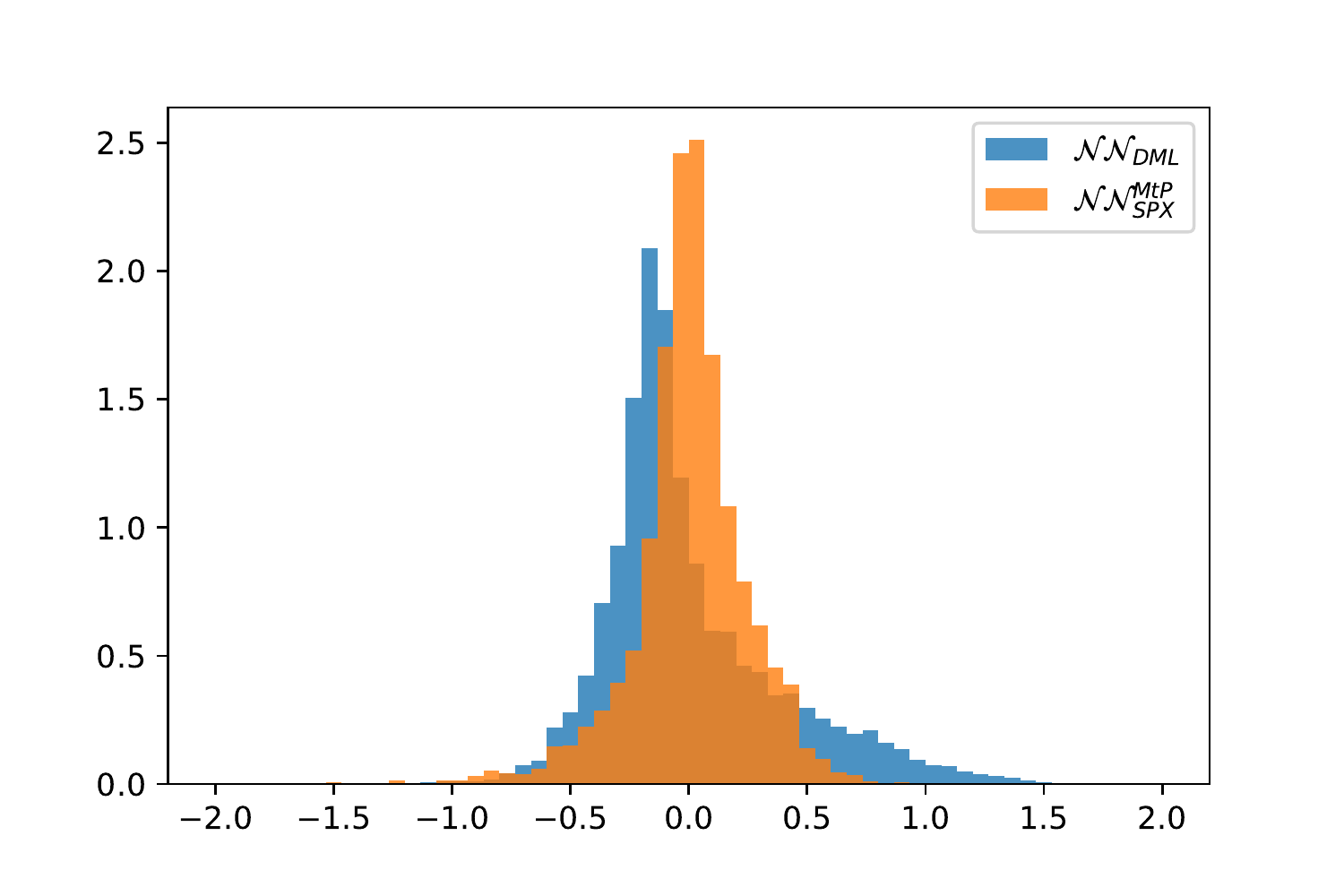}
    \caption{$\Delta_t = 0.0012$ (three times daily)}
  \end{subfigure}%
  \begin{subfigure}{.5\textwidth}
    \centering
    \includegraphics[width=\linewidth]{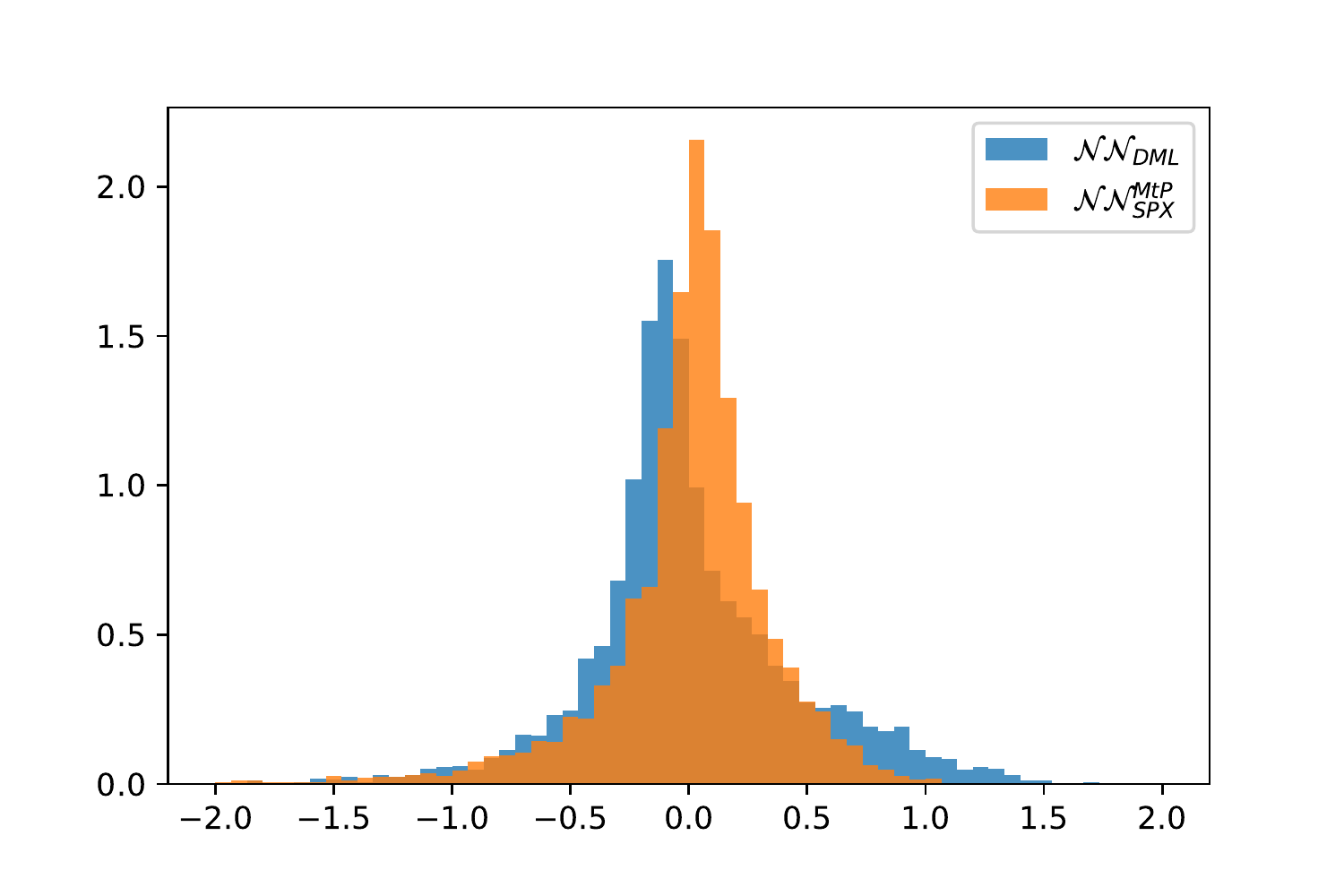}
    \caption{$\Delta_t = 0.0036$(daily)}
  \end{subfigure}
  \caption{Hedging P\&L with $\mathcal{NN}_{DML}$ and $\mathcal{NN}_{SPX}^{MtP}$ on simulated data. 
          $P_0$ given by Monte-Carlo is 2.9.}
  \label{fig:aad_nn_hedge_res}
\end{figure}

\textbf{Hedging on synthetic data}
\vskip 0.15in
\noindent
Without loss of generality, we take the following parameters in the experiments:
\begin{itemize}
  \item $\boldsymbol{\omega} = (1, 1.2, 0.35, 0.2, 0.0025)$
  \item $S_0 = 100, Z^i_0=0, i=1,\cdots,10$
  \item $K=98, T=0.08$
\end{itemize}
\noindent First we generate 50,000 paths of $(S_t)_{0\leq t\leq T}$ by following the scheme \ref{eq:euler_scheme} with time step $\Delta t=0.0012$. 
The price $P_0$ is estimated by the average of pathwise payoffs. Then we evaluate $\mathcal{J}_T$ for 5000 paths among them, with hedging time step
$\Delta t \in \{0.0012, 0.0036\}$.
From Figure \ref{fig:aad_nn_hedge_res} we can see that both methods lead to hedging payoffs around 0. Hedging less frequently brings slightly larger
variance of payoffs. We also remark that the method with $\mathcal{NN}^{MtP}_{SPX}$ generates smaller variance than the other with $\mathcal{NN}_{DML}$.
It is expected as the latter is trained with pathwise labels while the former is trained with ``true'' labels given by Monte-Carlo means, 
which have certainly smaller variance.

\noindent

\vskip 0.2in
\noindent 
\textbf{Hedging on market data}
\vskip 0.15in
\noindent
We perform daily hedging on two SPX monthly calls:
\begin{enumerate}
  \item Maturity date 16 June 2017, strike 2425, hedged since 18 May 2017.
  \item Maturity date 16 February 2018, strike 2750, hedged since 18 January 2018.
\end{enumerate}
\begin{figure}[!h]
  \centering
  \begin{subfigure}{.5\textwidth}
      \centering
      \includegraphics[width=\linewidth]{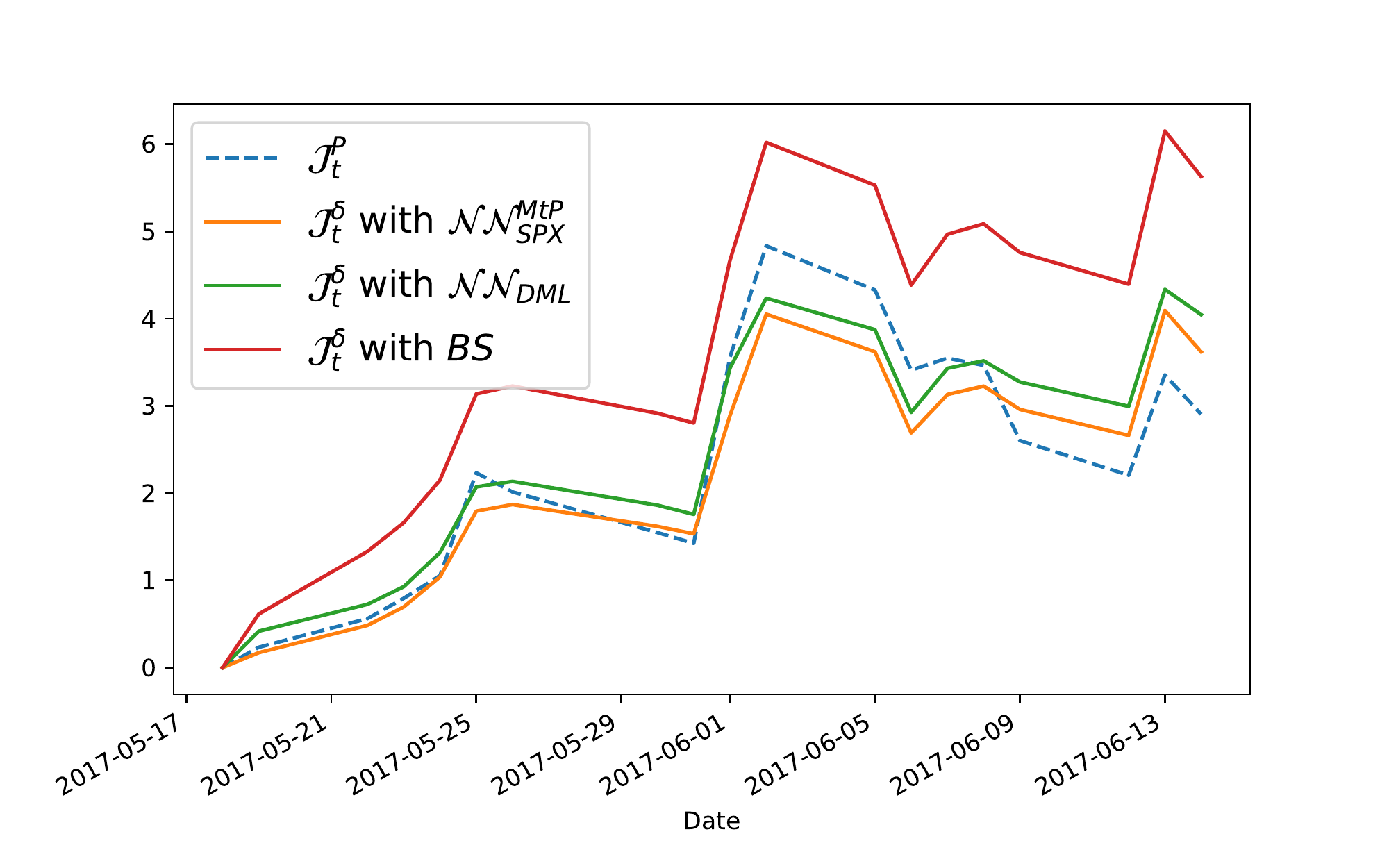}
  \end{subfigure}%
  \begin{subfigure}{.5\textwidth}
    \centering
    \includegraphics[width=\linewidth]{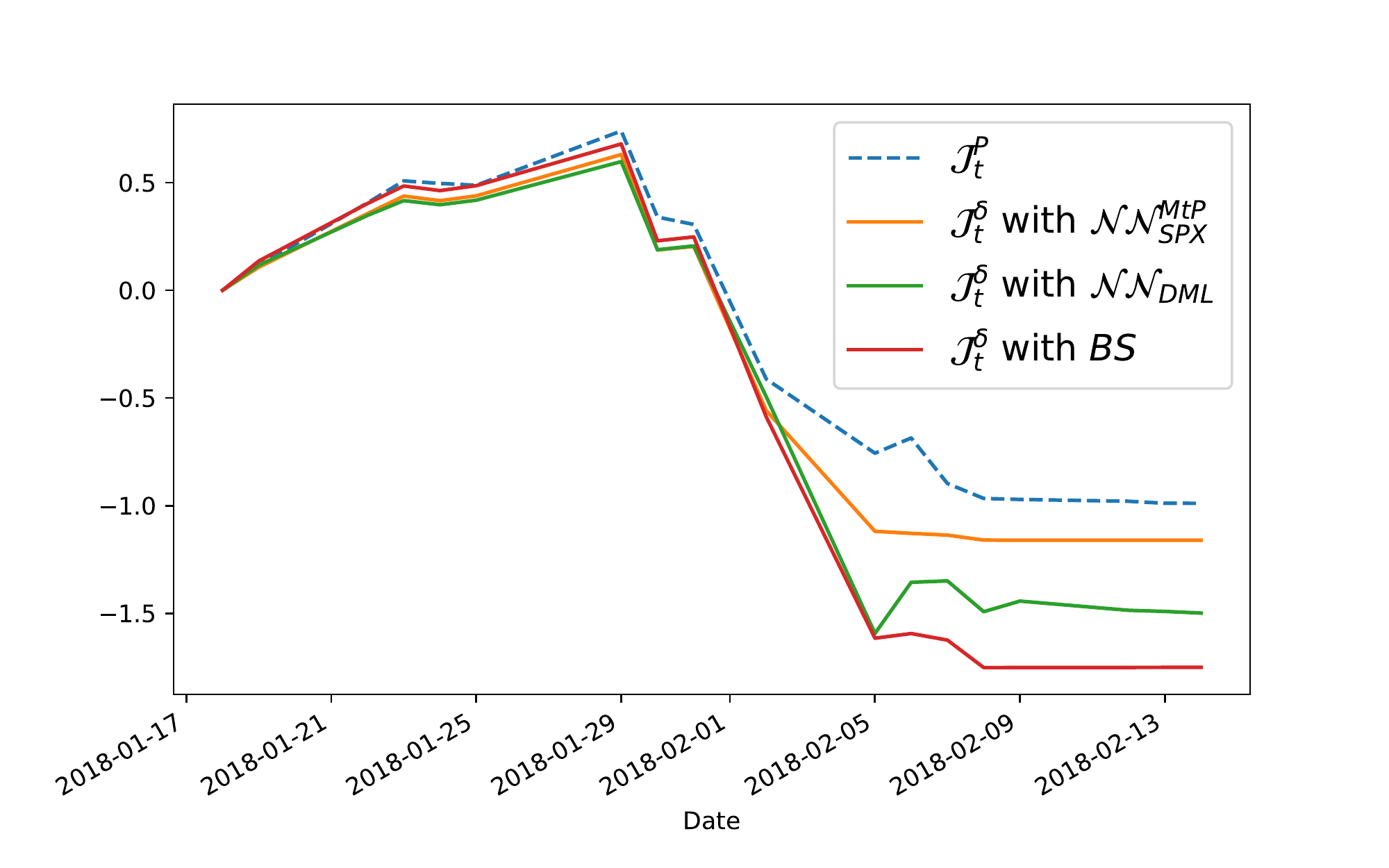}
  \end{subfigure}
  \caption{Hedging P\&L on two SPX calls with maturity 16 June 2017, strike 2425 (left), and maturity 16 February 2018, strike 2750 (right).}
  \label{fig:spx_hedge_market}
\end{figure}

\noindent On first day of hedging, we take market data for model calibration and get $\boldsymbol{\omega}$ and $\mathbf{z}_0$.
$\mathcal{NN}_{DML}$ is then trained on paths generated under $\boldsymbol{\omega}$.
Then for each following day, we update the value of factors $(Z^i)_{i=1, \cdots,10}$ with respect to the evolution of SPX, 
and we compute the hedging portfolio with $\mathcal{NN}_{SPX}^{MtP}$ and $\mathcal{NN}_{DML}$ as explained in the above.
Besides, we also test with Black-Scholes model where the implied volatility of the first day is used to compute \textit{Delta} as hedging ratio. 
Figure \ref{fig:spx_hedge_market} presents the evolution of $\mathcal{J}^P$ and $\mathcal{J}^{\delta}$ for these two examples.
Note that all quantities are divided by $P_0$ to be unitless. We see that $\mathcal{NN}_{SPX}^{MtP}$ and $\mathcal{NN}_{DML}$ 
can follow very well the market price of options, with smaller $|\mathcal{J}_T|$ than Black-Scholes approach. 
Of course, in practice we need to consider more elements like hedging cost, slippage, and to do more tests for systematic comparison, but
this is out of the scope of our current work.

\section{Conclusion}
We have seen that the deep neural networks can be used in calibrating the quadratic rough Heston model with reliable results.
The training of network demands indeed lots of simulated data, especially when the dimension of model parameters is high.
However, it is done off-line only once and the network will be reusable in many situations. 
Under the particular setting of our model, we can also use the network for risk hedging. Certainly, we can still improve the results
presented in the above, for example fixing finer grids of strikes and maturities, or using more factors for the approximation. 
We emphasize that the methodologies presented in our work are of course not limited to the model introduced here, and can be adapted to other models and other 
financial products. 

\newpage
\begin{appendices}
  \section{Kernel function approximation and simulation scheme}
  \label{ap:kernel_aprox}
\begin{figure}[!h]
  \centering
    \includegraphics[width=.7\linewidth]{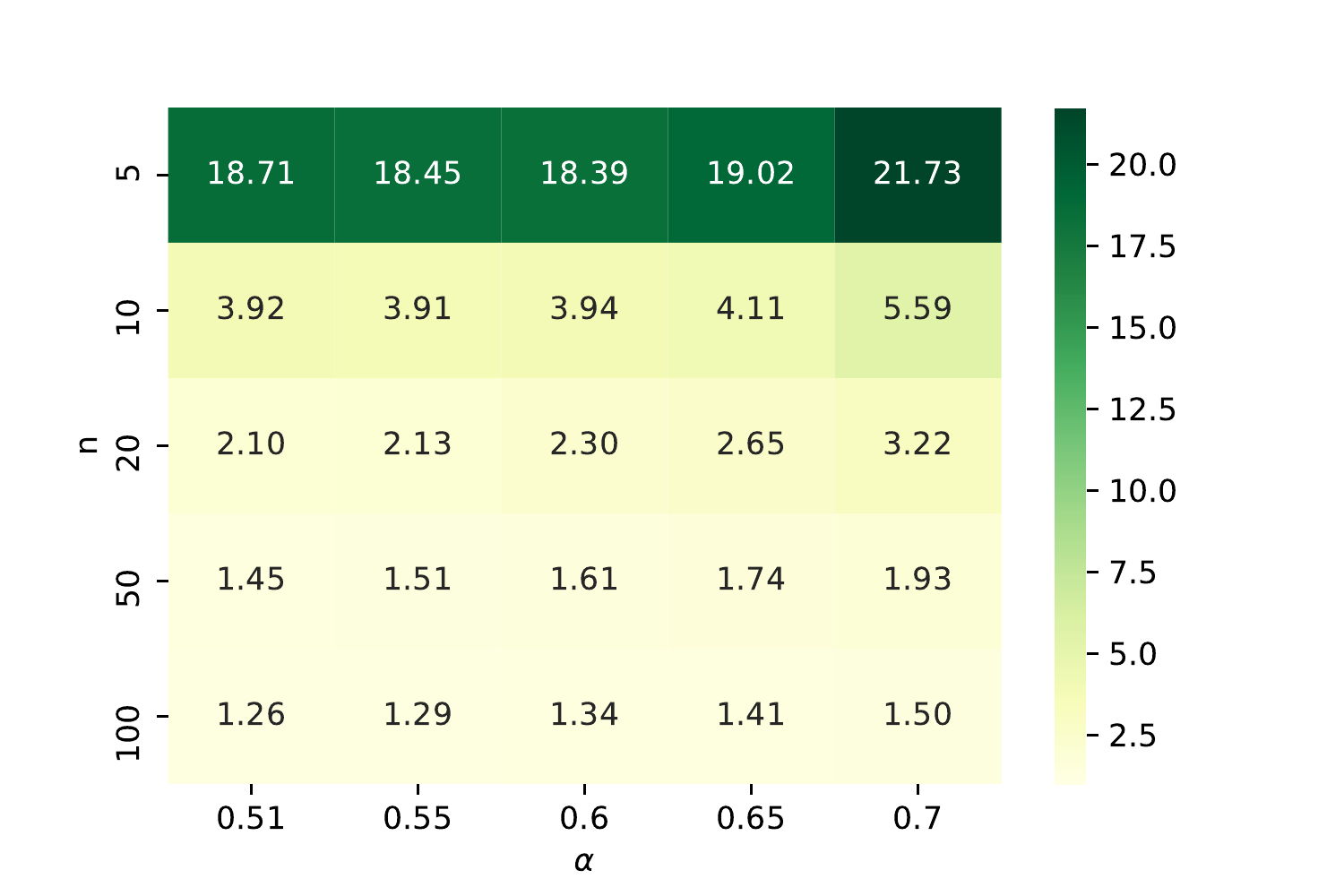}
    \caption{$x^*_n$ for different n and $\alpha$.}
    \label{fig:optimal_x}
\end{figure}

\begin{figure}
  \centering
  \begin{minipage}{.49\textwidth}
    \centering
    \includegraphics[width=\linewidth]{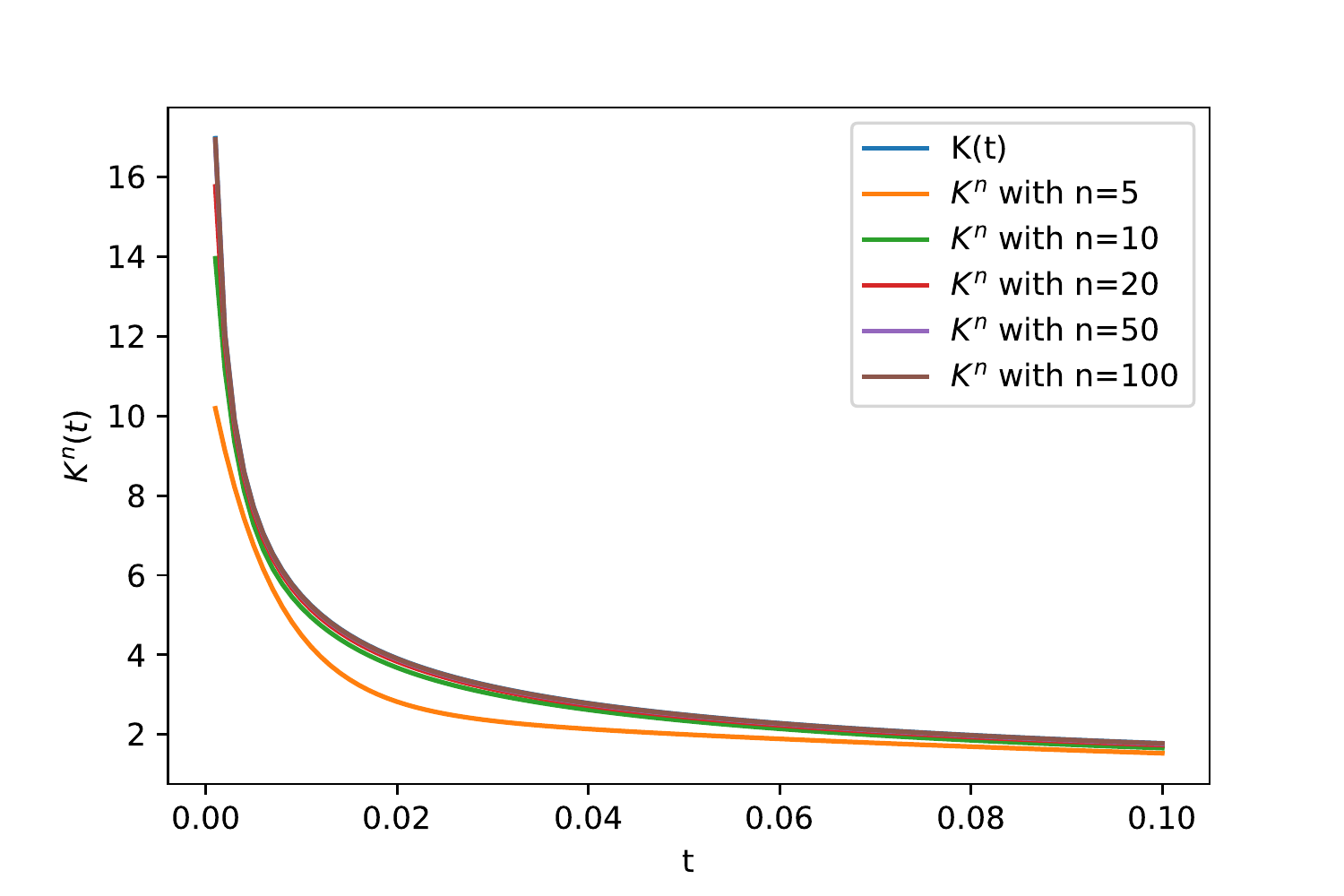}
    \captionof{figure}{$K(t)$ and $K^n(t)$ with different $n$ with $\alpha=0.51$.}
    \label{fig:K_t}
  \end{minipage}%
  \begin{minipage}{.49\textwidth}
    \centering
    \includegraphics[width=\linewidth]{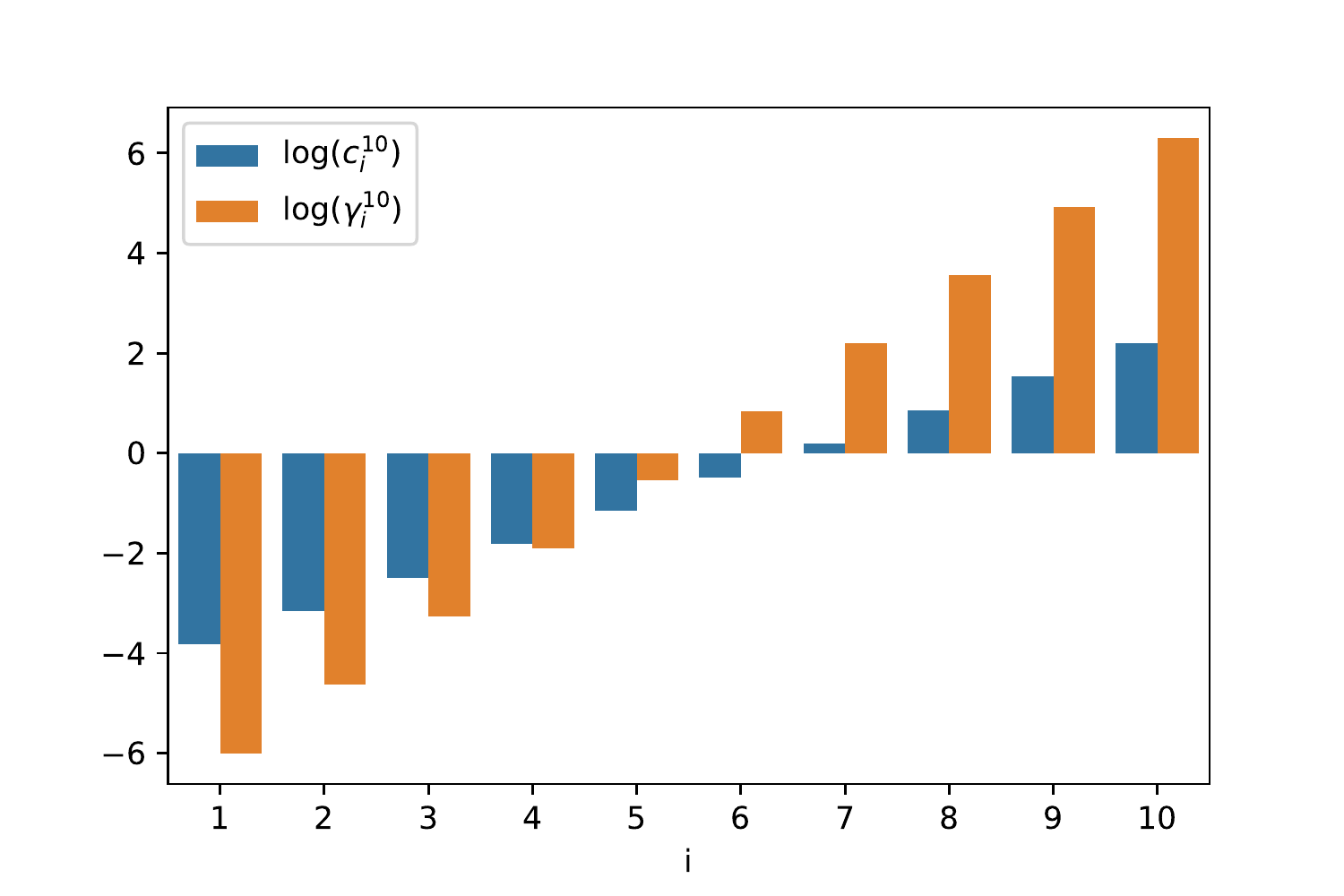}
    \captionof{figure}{Logarithm of $c_i^{10}$ and $\gamma_i^{10}$ by taking $x_{10}^{*}=3.92$.}
    \label{fig:coefs_bar}
  \end{minipage}
\end{figure}

  Here we recall the geometric partition of $(c^n_i, \gamma^n_i)_{i=1,\cdots, n}$ proposed in \cite{abi2019lifting}:
  $$
    c^n_i = \int_{\eta^n_{i-1}}^{\eta^n_i}\mu(dx), \quad \gamma^n_i=\frac{1}{c^n_i}\int_{\eta^n_{i-1}}^{\eta^n_i}x\mu(dx), \quad i=1,\cdots,n \, ,
  $$
  with $\mu(dx)=\frac{x^{-\alpha}}{\Gamma(\alpha)\Gamma(1-\alpha)}dx$ and $\eta^n_i = x^{i-n/2}_n$. Hence we have
  $$
    c^n_i = \frac{x_n^{(1-\alpha)(i-n/2)}(1-x_n^{\alpha-1})}{(1-\alpha)\Gamma(\alpha)\Gamma(1-\alpha)}, \quad \gamma^n_i = \frac{(1-\alpha)x_n^{i-1-n/2}(x_n^{2-\alpha}-1)}{(2-\alpha)(x_n^{1-\alpha} - 1)} \, .
  $$
\noindent
Given $n$, $\alpha$ and $T$, we can determine the ``optimal'' $x_n$ as
$$
  x^{*}_n(\alpha, T) = \arg \min_{x^*_n>1}\|K^n - K\|_{L^2(0, T)} \, .
$$
\noindent We fix $T=0.1$ as we are more interested in short maturities and we get the optimal $x^*_n$ for different $n$ and $\alpha$ as shown in Figure \ref{fig:optimal_x}. 
It is consistent with the analysis in \cite{abi2019lifting} that given $\alpha$, we need to increase $x_n$ to mimic roughness with less factors. Given $n$, $x^*_n$ does not change
a lot with $\alpha$, which indicates that in practice we can actually fix $x^*_n$ independently of $\alpha$.
\noindent
Choosing a good $n$ is a trade-off between simulation efficiency and good approximation of rough volatility models.
In our test, we take $n=10$ and $x_{10}^*=3.92$. On one hand, we can see from Figure \ref{fig:K_t} that 
the approximation of $K^{10}$ to $K$ is not far away from other $K^n$ with larger $n$. 
On the other hand, it means a margin for improvement with larger $n$. Figure \ref{fig:coefs_bar} gives the resulting $c_i^n$
and $\gamma_i^n$. We see that the wide range of $\gamma$ contributes to the multi-timescales nature of volatility processes.

\noindent
We discard the notation $n$ and use the following modified explicit-implicit Euler scheme for simulating Model (\ref{eq:V_n}-\ref{eq:Z_factor}):
\begin{equation}
    \begin{split}
      \hat{S}_{k+1} &= \hat{S}_k + \hat{S}_k\sqrt{\hat{V}_k}(W_{k+1} - W_k), \qquad \hat{V}_k = a(\hat{Z}_k - b)^2 + c \, ,\\ 
      \hat{Z}^{i}_{k+1} &= \frac{1}{1+\gamma_i\Delta t}\big(\hat{Z}^{i}_k-\lambda\hat{Z}_k\Delta t + \eta\sqrt{\hat{V}_k}(W_{k+1} - W_k)\big) \, , \\
      \hat{Z}_k &= \sum_{i=1}^nc_i\hat{Z}^{i}_k \, ,
    \end{split}
    \label{eq:euler_scheme}
\end{equation}

\noindent
for a time step $\Delta t=T/N$, $k=1,\cdots, N$ and $(W_{k+1} - W_k) \sim \mathcal{N}(0, \Delta t)$. 
One could also use the explicit sheme for $Z^i$ given by
$$
  \hat{Z}^{i}_{k+1} = (1-\gamma_i\Delta t)\hat{Z}^{i}_k-\lambda\hat{Z}_k\Delta t + \eta\sqrt{\hat{V}_k}(W_{k+1} - W_k) \, .
$$
However, with above $x_{10}^*$, we get $\gamma_{10}=542.32$. One then need $\Delta t $ to be necessarily small to ensure the scheme's stability.
Instead we could use 
$$
  \hat{Z}^{i}_{k+1} - \hat{Z}^{i}_k = -\gamma_i\Delta t\hat{Z}^{i}_{k+1} - \lambda\hat{Z}_k\Delta t + \eta\sqrt{\hat{V}_k}(W_{k+1} - W_k) \, ,
$$
which leads to Scheme \ref{eq:euler_scheme} and avoids this issue. 

  \section{Network training with transfer learning}
  \begin{figure}[!h]
    \centering
    \includegraphics[width=\textwidth]{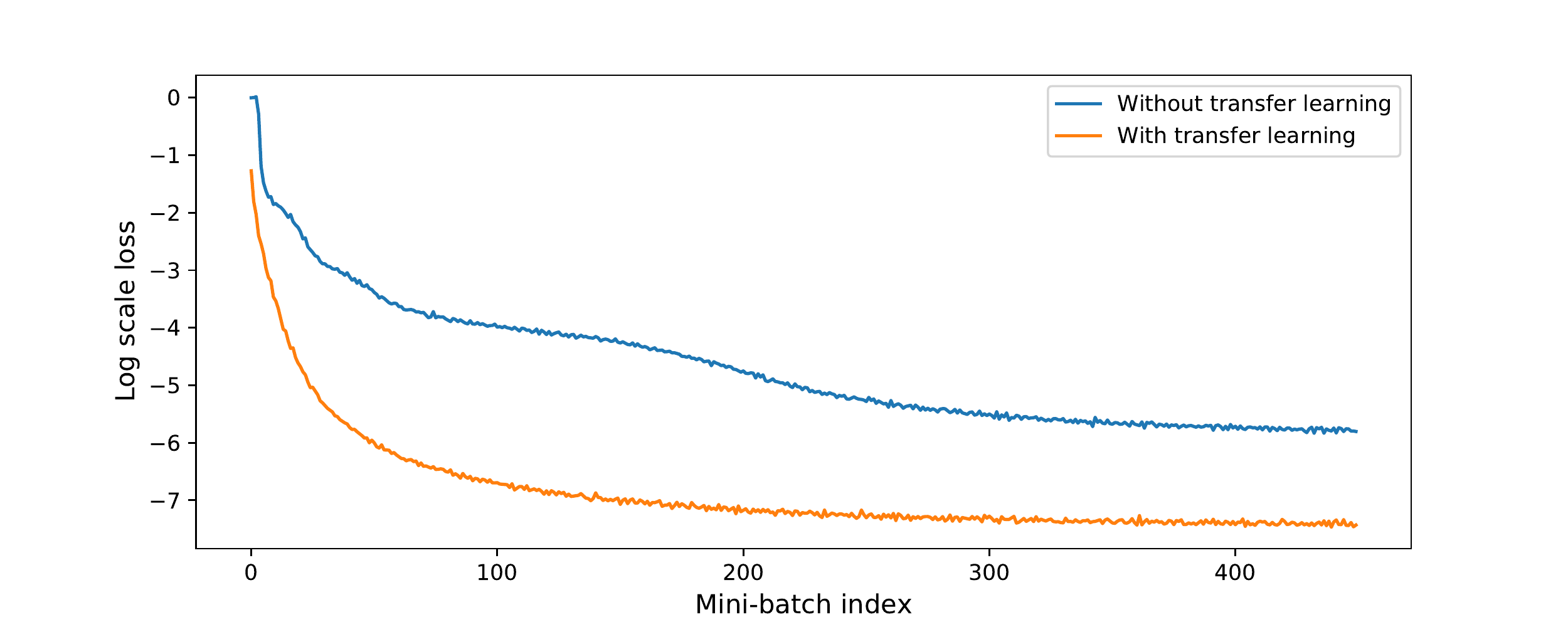}
    \caption{Training loss evolution for $\alpha=0.65$ with 10,000 samples.}
    \label{fig:t_learn_loss}
  \end{figure}
  \label{ap:transfer}
  When we switch to models with $\alpha$ not equal to $0.51$, we can apply the idea of transfer learning to accelerate network training. 
  More precisely, we use the parameters of the network corresponding to the case $\alpha=0.51$ to initialise the network for cases with different $\alpha$. 
  Here we give an example on the training of $\mathcal{NN}_{SPX}^{MtP}$ with $\alpha=0.65$. With 10,000 training samples, we can see from Figure \ref{fig:t_learn_loss} that transfer learning can 
  help the training converge much faster to a lower loss than the one with random parameter initialization.

  \section{Pricing and calibration with neural networks}
  \label{sec:rel_err}
  \subsection{On simulated data}
\begin{figure}[!h]
    \centering
    \includegraphics[width=\textwidth]{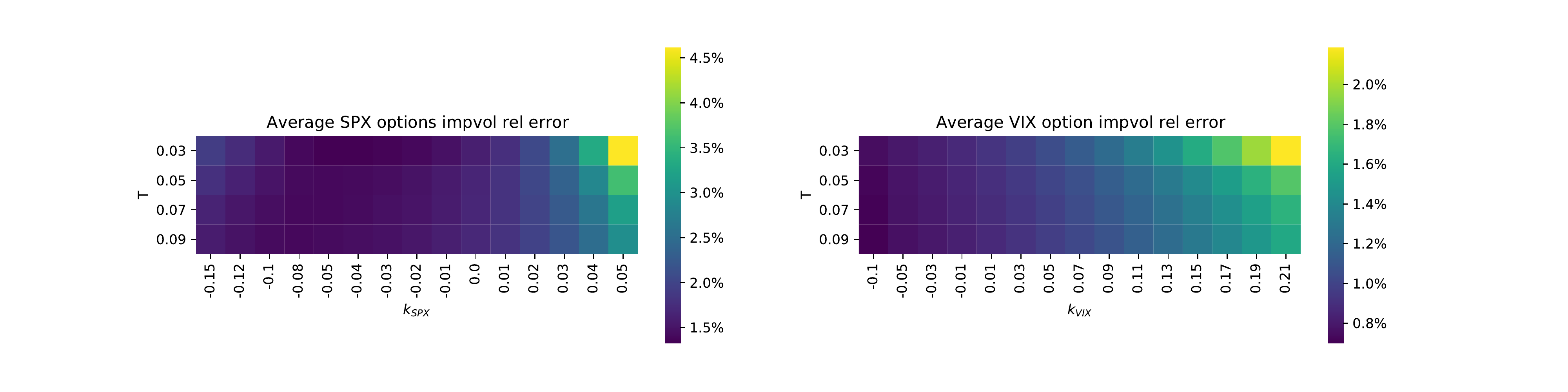}
    \caption{Average Monte-Carlo relative errors of implied volatilities across test set, defined as the normalized half 95\% confidence
    interval of Monte-Carlo simulations.}
    \label{fig:mc_spx_vix_rel_err}
  \end{figure}

\begin{figure}[!h]
    \centering
    \includegraphics[width=\textwidth]{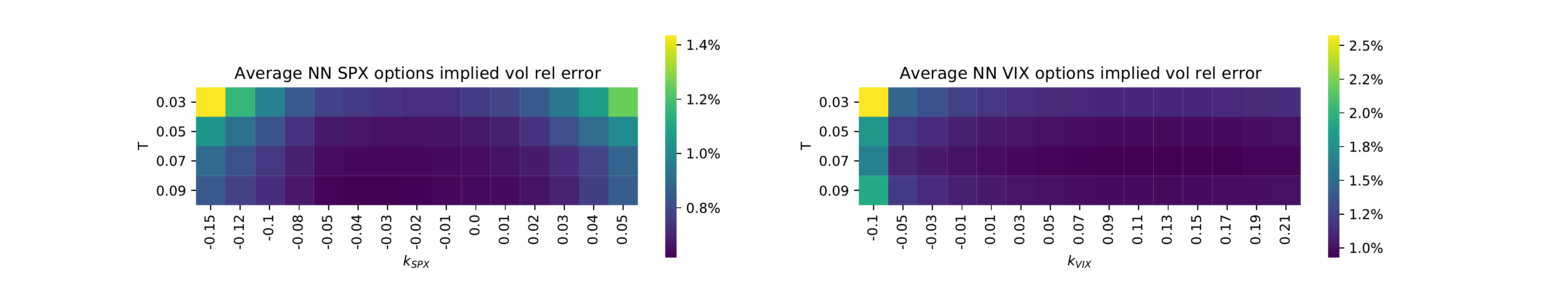}
    \caption{Average relative errors of implied volatilities across test set for $\mathcal{NN}^{MtP}_{SPX}$ and $\mathcal{NN}^{MtP}_{VIX}$.}
    \label{fig:nn_spx_vix_rel_err}
  \end{figure}

\begin{figure}[!h]
    \centering
    \includegraphics[width=\textwidth]{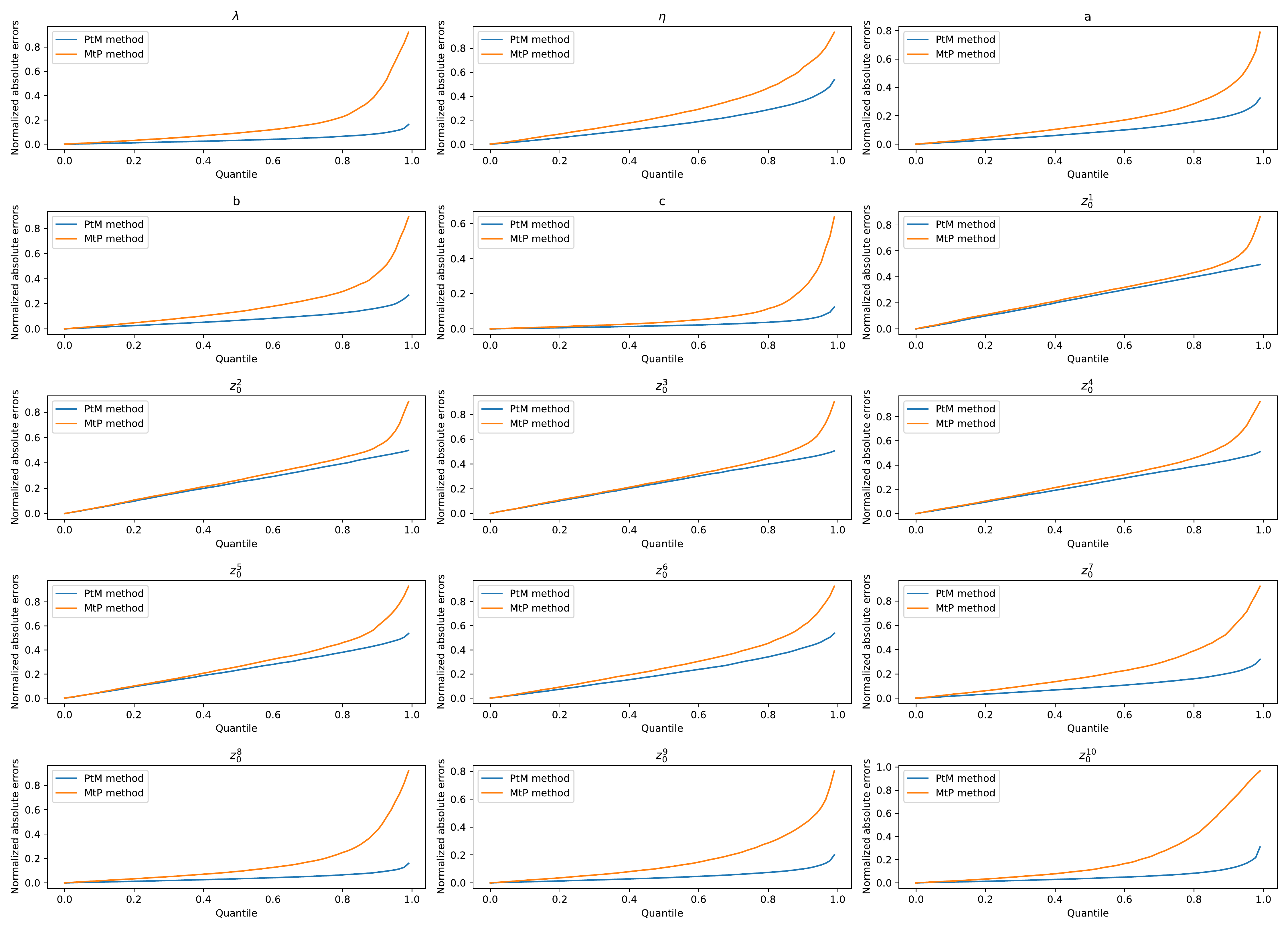}
    \caption{Empirical CDF of NAE of calibrated parameters.}
    \label{fig:nn_calib_err}
  \end{figure}

  Here we present the average relative pricing errors across test as an alternative evaluation metric. Figure \ref{fig:mc_spx_vix_rel_err} stands for
  the benchmark given by Monte-Carlo and Figure \ref{fig:nn_spx_vix_rel_err} gives the results with $\mathcal{NN}_{SPX}^{MtP}$ and $\mathcal{NN}_{VIX}^{MtP}$.
  Figure \ref{fig:nn_calib_err} shows the accuracy of calibrated parameters by plotting the empirical CDFs of NAE. 

  \newpage
  \subsection{On market data}
  \label{app:calib_sample_1}
\begin{figure}[!h]
    \centering
    \includegraphics[width=\textwidth]{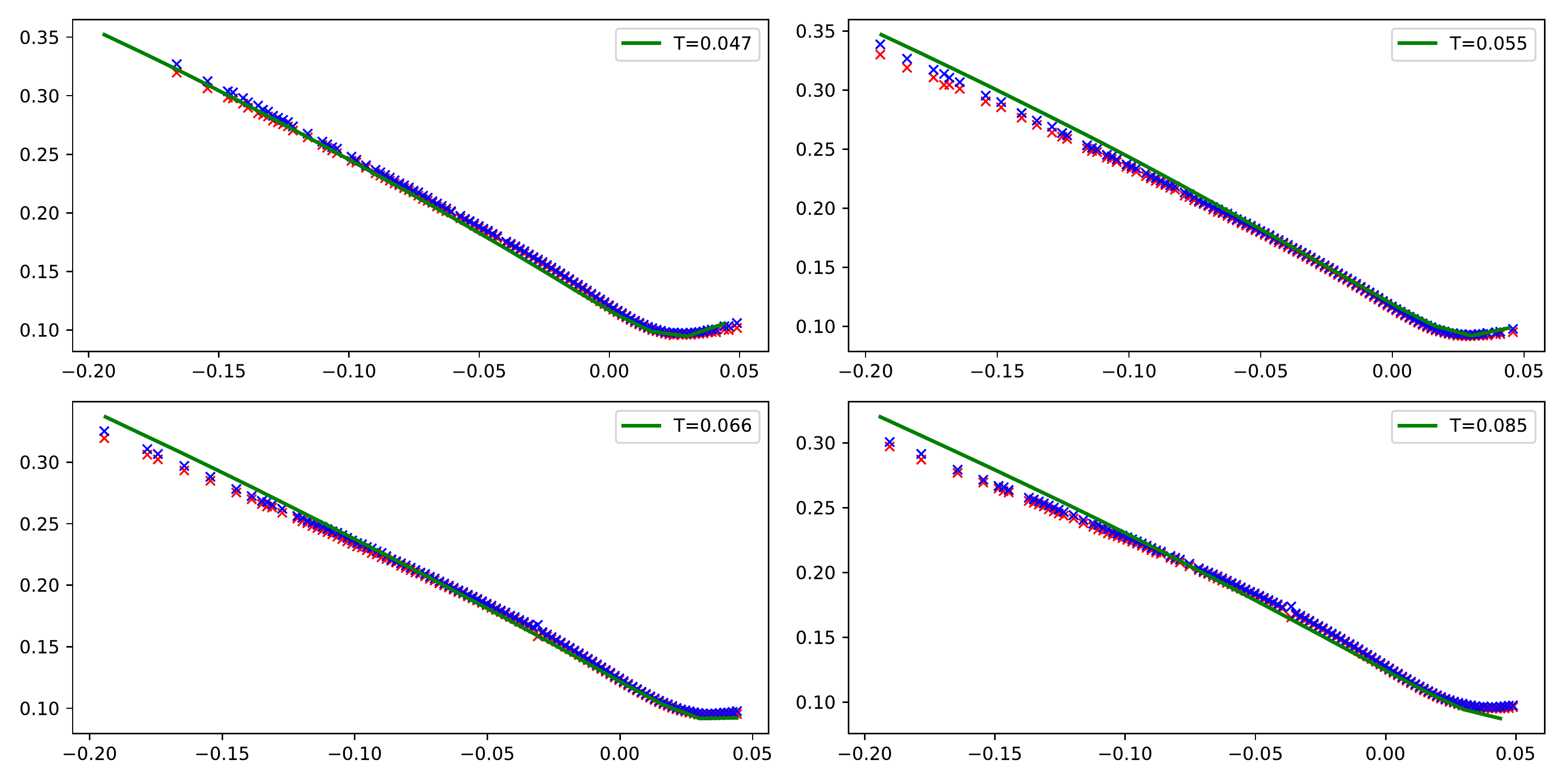}
    \caption{Implied volatilities fit on SPX options for 10 September 2019. Bid and ask of market volatilities are represented respectively 
    by red and blue points. Green line is the output of model with Monte-Carlo method.}
    \label{fig:nn_spx_calib_sample_1}
  \end{figure}

\begin{figure}[!h]
    \centering
    \includegraphics[width=\textwidth]{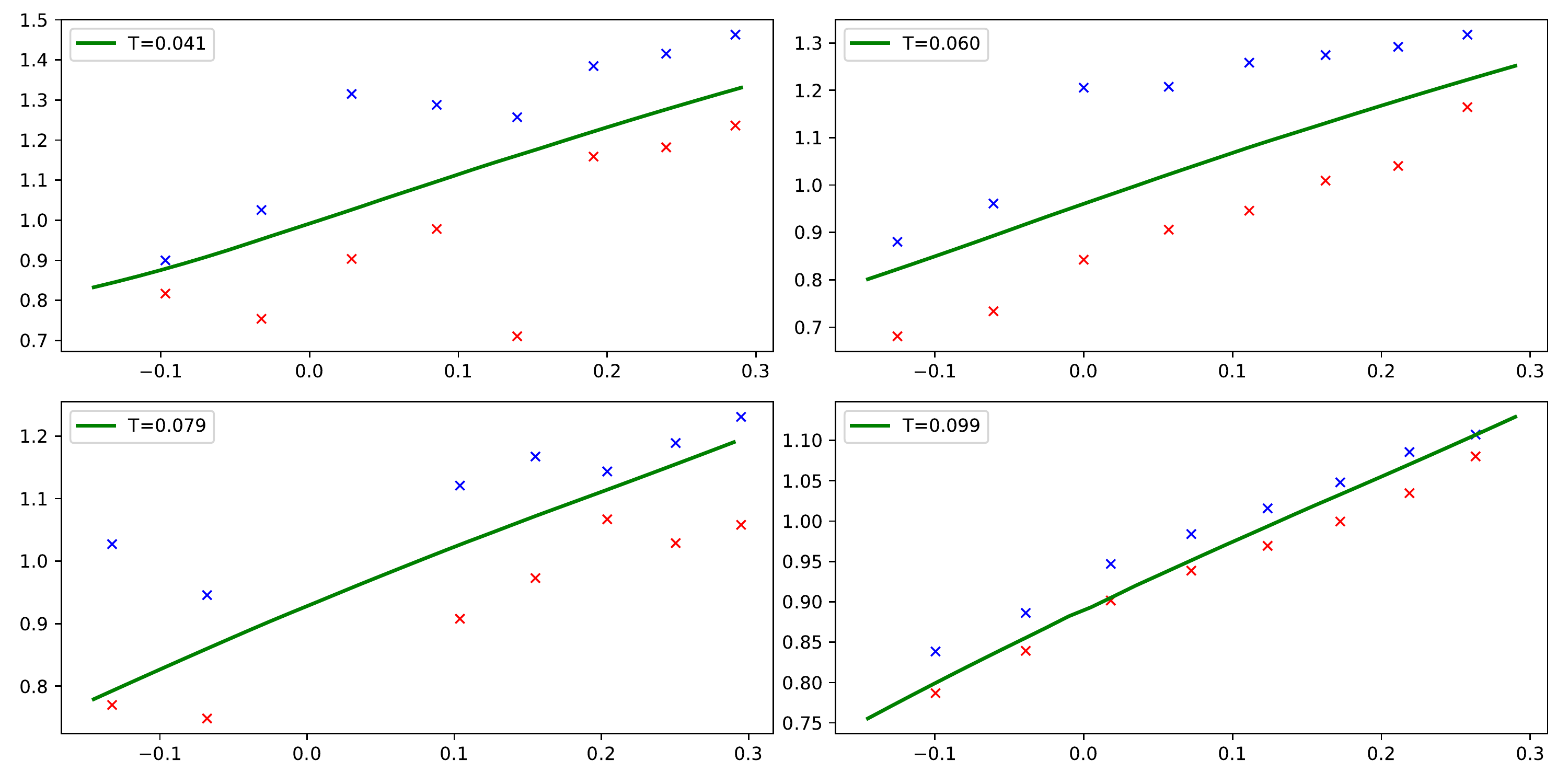}
    \caption{Implied volatilities fit on VIX options for 10 September 2019. Bid and ask of market volatilities are represented respectively 
    by red and blue points. Green line is the output of model with Monte-Carlo method.}
    \label{fig:nn_vix_calib_sample_1}
  \end{figure}

\begin{figure}[!h]
    \centering
    \includegraphics[width=0.8\textwidth]{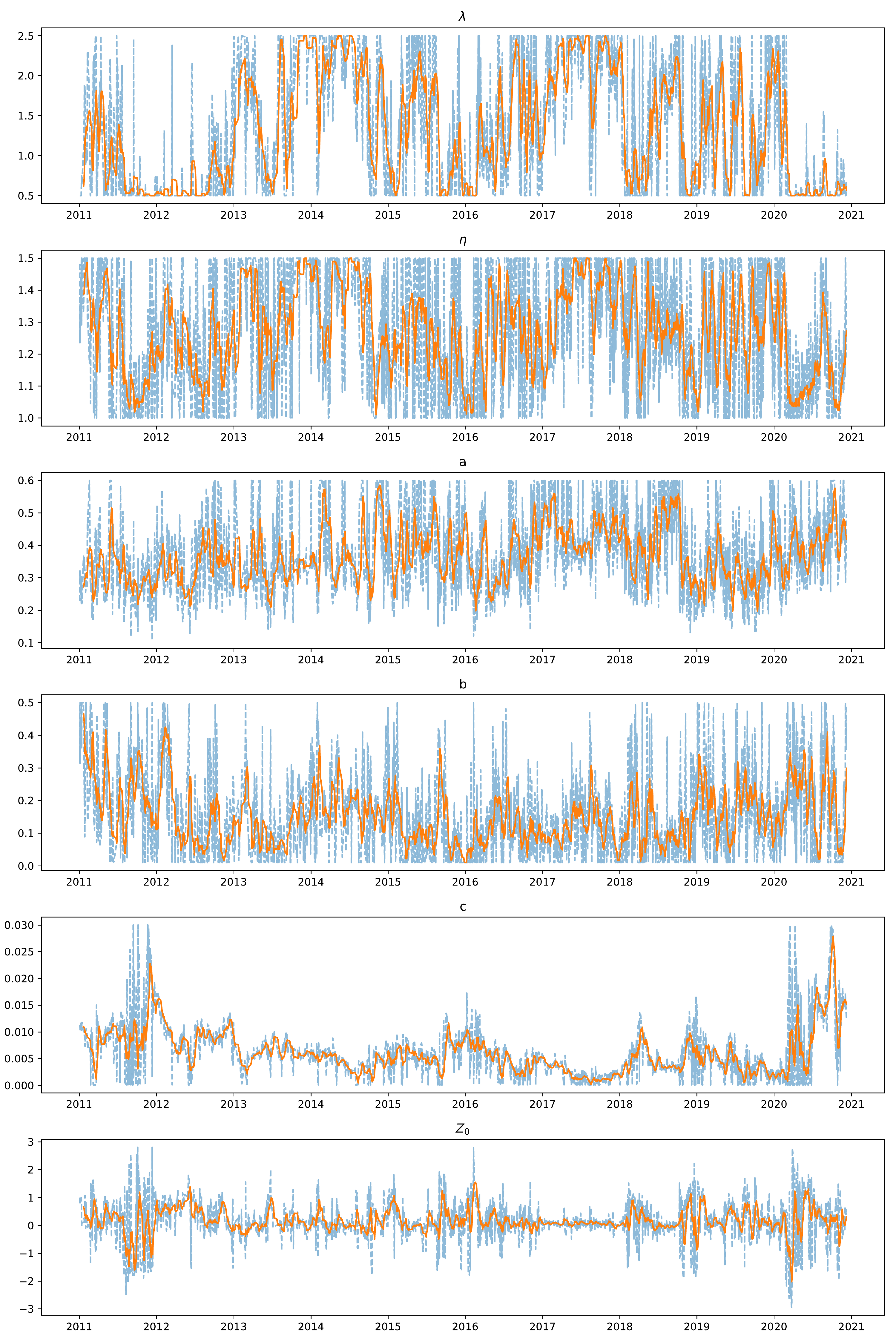}
    \caption{Historical dynamics of calibrated parameters. The orange lines are 10-day moving average of parameters. Recall that $Z_0 := \sum_{i=1}^{10}c_iz^{i}_0$.}
    \label{fig:market_calib_params}
  \end{figure}

  We give another IVS fit example on 10 September 2019. The parameters calibrated are:
  \begin{align*}
    \boldsymbol{\omega} &= (1.717, 1.5, 0.265, 0.246, 0.0001) \, ,\\
    \mathbf{z}_0 &= (-0.009, 0.015, 0.011, 0.036, 0.002, -0.011, -0.018, 0.074, 0.142, -0.171) \, . 
  \end{align*}
  The Monte-Carlo results with these parameters are shown in Figure \ref{fig:nn_spx_calib_sample_1} and Figure \ref{fig:nn_vix_calib_sample_1}.
  Figure \ref{fig:market_calib_params} presents the historical dynamics of parameters from daily calibration on market data.

\end{appendices}

\clearpage
\bibliography{ref}
\bibliographystyle{abbrv}
\end{document}